\documentclass[aps,pre,reprint,longbibliography]{revtex4-1}
\usepackage{graphicx}
\usepackage{amsmath}
\usepackage{amssymb}
\usepackage{natbib}
\usepackage[usenames]{color}

\newcommand{\change}{}

\begin{document}

\title{ 
Exact solutions and physical analogies for unidirectional flows
}

\author{Martin Z. Bazant 
}

\affiliation{Departments of Chemical Engineering and Mathematics,
  Massachusetts Institute of Technology, Cambridge, MA 02139 USA}
\affiliation{ \protect{\em Present address:} Department of Materials Science and Engineering and SUNCAT Interfacial Science and Catalysis, Stanford University, Stanford, CA 94305}
\setcounter{page}{1} \date{\today}
\label{firstpage}

\begin{abstract}
Unidirectional flow is the simplest phenomenon of fluid mechanics.  Its mathematical description, the Dirichlet problem for Poisson's equation in two dimensions with constant forcing, arises in many physical contexts, such as the torsion of elastic beams, first solved by de Saint-Venant for complex shapes.  Here, the literature is unified and extended by identifying \change{seventeen} physical analogies for unidirectional flow and describing their common mathematical structure.  Besides classical analogies in fluid and solid mechanics, applications are discussed in stochastic processes (first passage in two dimensions), pattern formation (river growth by erosion), and electrokinetics (ion transport in nanochannels), which also involve Poisson's equation with non-constant forcing.   Methods are given to construct approximate geometries that admit exact solutions, by adding harmonic functions to quadratic forms or by truncating eigenfunction expansions.  \change{ Exact solutions for given geometries are also derived by conformal mapping. } We prove that the remarkable geometrical interpretation of Poiseuille flow in an equilateral triangular pipe  (the product of the distances from an interior point to the sides) is only shared by parallel plates and unbounded equilateral wedges (with the third side hidden behind the apex).    We also prove Onsager reciprocity for linear electrokinetic phenomena in straight pores of arbitrary shape and surface charge, based on the mathematics of unidirectional flow. 
\end{abstract}

\maketitle

\begin{figure*}
\begin{center}
\vspace{0.2in}
\includegraphics[width=\linewidth]{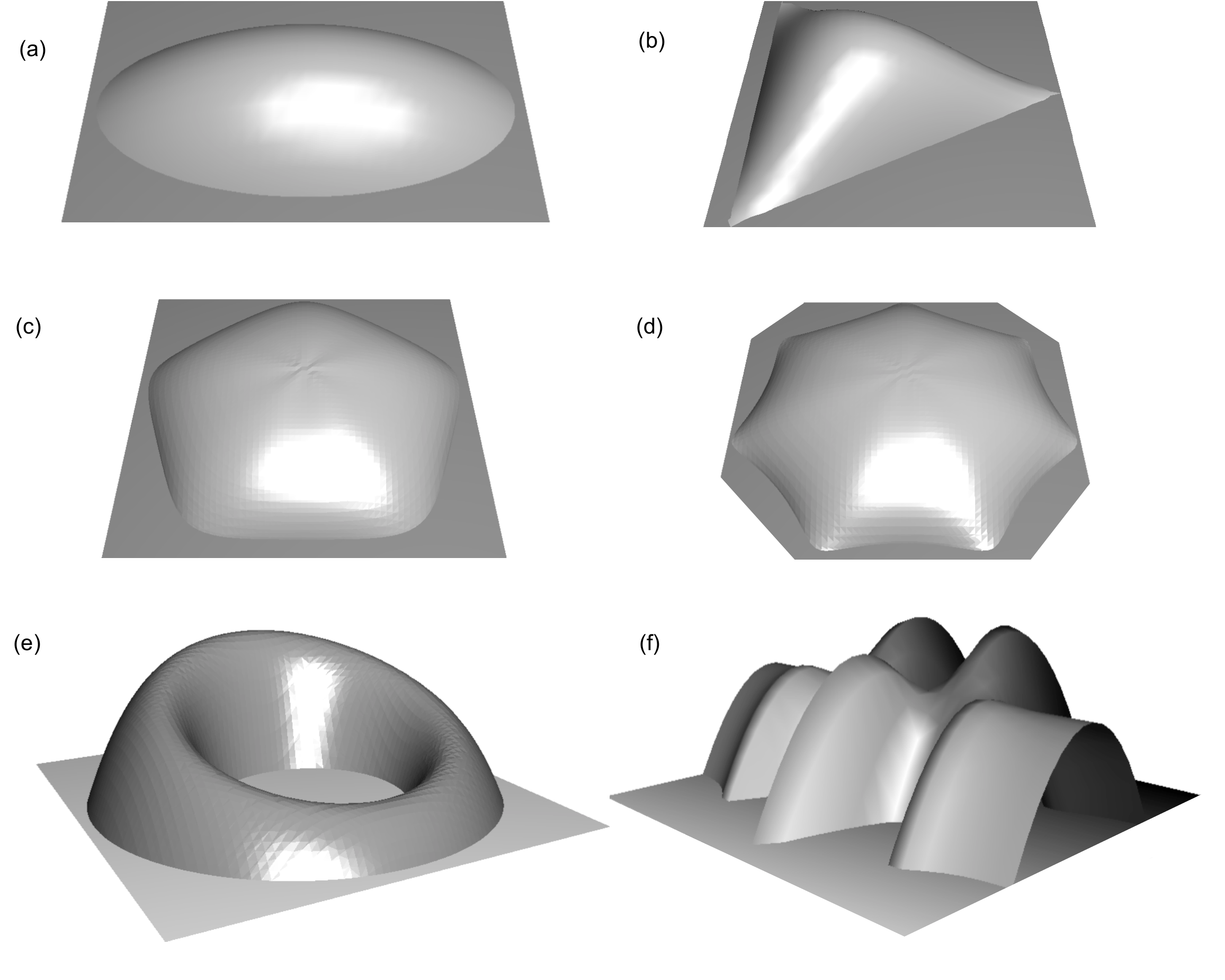}  
\caption{ Exact solutions for unidirectional (Poiseuille) flow profiles in pipes of different cross sections, rendered as deformed membranes or soap bubbles.  The same solutions are derived and contour-plotted below:  (a) ellipse (Fig. ~\ref{fig:ell}(a)), (b) equilateral triangle (Fig.~\ref{fig:tri}), (c) rounded pentagon (Fig. ~\ref{fig:rounded}(a)), (d) seven-pronged star (Fig. ~\ref{fig:rounded})(e), (e) off-center coaxial pipes (Fig.~\ref{fig:coax}(b)),  and (f) grooved parallel plates (Fig.~\ref{fig:twist}(c)).
\label{fig:bubbles}}
\end{center}
\end{figure*}

\section{ Introduction }

Precious few exact solutions of the Navier-Stokes equations are known, but they serve to guide our thinking about fluid mechanics \citep{wang1991,wang1989,bazant2005}.  Most students first encounter the parabolic profile of Poiseuille flow in a circular pipe or between flat plates~\citep{sutera1993}.  Some less familiar examples are shown in Fig. \ref{fig:bubbles}.   In these unidirectional flows, inertia plays no role, leading to a simple balance between viscous stress and the applied pressure gradient.  

\change{ Poiseuille's law for the flow rate in a narrow capillary~\cite{sutera1993,poiseuille1847} was apparently first derived from the Navier-Stokes equations by Stokes himself~\cite{stokes1845} in 1845.  He was uncertain about boundary conditions and included an unknown slip velocity, later calculated by \citet{butcher1876}, building on ideas of \citet{navier1823}.  The derivation without slip is normally attributed to \citet{hagenbach1860} and \citet{jacobson1860} in 1860, although a decade earlier, Stokes had already used the no-slip boundary condition in his famous paper on viscous drag~\cite{stokes1851}. 
}

\change{
In the original paper on viscous flow, Stokes remarked that it is ``extremely easy" to derive the velocity profile in a circular capillary~\cite{stokes1845}, but he surely appreciated the challenges posed by other geometries.  Shape dependence was first analyzed by \citet{venant1855} in 1855 in the seemingly different context of torsion of elastic beams.  In 1871, \citet{bous1871} recognized that de Saint-Venant's theory of torsion is mathematically equivalent to Stokes' theory of Poiseuille flow.  Ten years later \citet{heaviside1881} noted the equivalence of beam torsion and the magnetic self-induction of an electrical wire, which was eventually recognized as another analogy for pipe flow~\cite{hunt1995}, along with the ``membrane analogy" identified by \citet{prandtl1903} in 1903. }

\change{
In the twentieth century, pipe flows for different cross-sectional shapes were extensively characterized in the engineering literature, especially for heat transfer in ducts~\citep{shah1978}, and the number of physical analogies also grew. 
}
The common mathematical problem involves Poisson's equation from electrostatics ~\cite{jackson_book},
\begin{equation}
- \nabla^2 u = k   \label{eq:flow}
\end{equation}
typically with constant forcing $k$ and Dirichlet (no-slip) boundary conditions on a two-dimensional domain.   The same problem arises in solid mechanics for beam torsion and bending~\citep{musk_book,soko_book,timo_book}, and myriad exact solutions have been derived.  Beginning with the seminal paper of \citet{venant1855},  many complex  shapes were analyzed by conformal mapping, \change{ notably by \citet{morris1939}}, \citet{musk_book}, and \citet{polya_book}.  \change{ Morris also applied her general solution for beam torsion to Poiseuille flow~\cite{morris1939}}, and \citet{tao1961,tao1962} later solved related problems in forced convection.   In contrast to the more familiar case of Laplace's equation, however, conformal mapping cannot be as easily applied to Poisson's equation, since it is not conformally invariant \citep{bazant2004}.

Despite this extensive literature, it is worth revisiting the mathematics of uniform flow in various modern contexts, such as microfluidics  \citep{kirby_book,tabeling_book}, \change{ transport in porous media~\cite{torquato_book}, } stochastic processes~\citep{redner_book,vankampen_book}, \change{ chemical reactions~\cite{calef1983}}, biological reactions~\citep{lauffenburger_book,berg_book},  architectural structures~\citep{pottman2007}, groundwater flow~\citep{dunne1980}, river growth~\citep{cohen2015}, and electrokinetic phenomena~\citep{hunter_book,bruus_book}, where approximate or numerical solutions have been used to treat complicated geometries.  The initial motivation for this work came from theoretical microfluidics \citep{bruus_book}.  \citet{mortensen2005} used boundary perturbation methods to approximate the hydraulic resistance of microchannels with  near-circular cross sections, and these approximations closely resemble exact solutions derived by \citet{venant1855}.  Similarly, eigenfunction expansions, such as the well known Fourier series for a rectangular cross section~\citep{bruus_book}, can also be viewed as sum of exact solutions, only for slightly different geometries. 

In this article, we develop the mathematics of unidirectional flow through a variety of examples, both old and new.  Although formal integral solutions can be derived for any geometry, we focus on the construction of special geometries that approximate domains of interest and admit simple exact solutions.    We begin by reviewing the history of this problem in fluid and solid mechanics and discussing many additional applications.

\begin{figure*}
\begin{center}
\vspace{0.2in}
\includegraphics[width=0.9\linewidth]{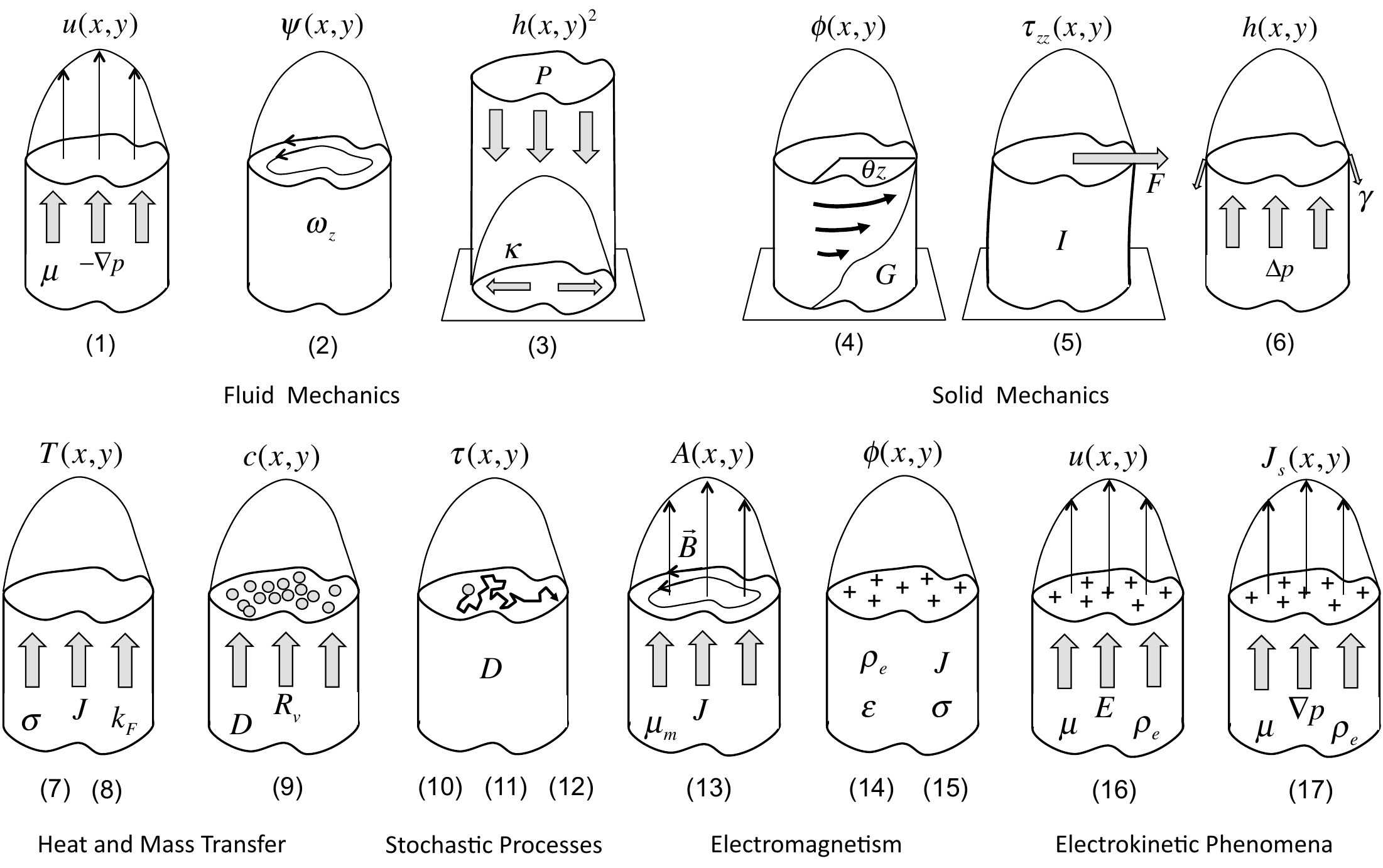}
\caption{ \change{ Seventeen} analogous physical phenomena from six broad fields, all described by Poisson's equation in two dimensions.  
{\it Fluid mechanics:} (1) Poiseuille flow in a pipe,  (2) circulating flow in tube of constant vorticity, and (3) groundwater flow fed by precipitation.  {\it Solid Mechanics:} (4) torsion or (5) bending of an elastic beam, and (6) deflection of a membrane, meniscus or soap bubble;  {\it Heat and mass transfer:}   
 (7) resistive heating of an electrical wire, (8) viscous dissipation in pipe flow and (9) reaction-diffusion in a catalyst rod; {\it Stochastic processes:}  (10) first passage time in two dimensions,  (11) the chain length profile of a grafted polymer in a tube, and (12) \change{ the mean rate of a diffusion-controlled reaction};  {\it Electromagnetism:}  \change{ (13) vector potential for magnetic induction in a shielded electrical wire}, and the electrostatic potential in (14) a charged cylinder or (15) a conducting sheet or porous electrode;  {\it Electrokinetic phenomena:} (16) electro-osmotic flow and (17) streaming current in a pore or nanochannel.
\label{fig:analogies} 
}
\end{center}
\end{figure*}

\section{ Physical Analogies }

Poisson's equation in two dimensions describes a remarkable variety of physical phenomena.  Exact solutions can be traced back (at least) to the seminal 1855 paper of \citet{venant1855} on the elastic deformation of straight, prismatic beams under torsion.  As noted above, the mathematical equivalence of beam torsion and pipe flow was first recognized by \citet{bous1871} and later extended to convective heat transfer (Marco and Han, 1955 \cite{marco1955}).    More easily visualized analogies are provided by the deflection of elastic membranes (Prandtl, 1903 \cite{prandtl1903}) or soap bubbles (Griffith, 1917 \citep{griffith1917}), and by the potential profile of electrically conducting sheets (Waner, 1953 \citep{waner1953}).  These mathematical insights allowed Poisson's equation to be solved {\it experimentally}, long before it could be solved numerically on a computer.

In this section, we survey the literature and expand the number of physical analogies to seventeen, sketched in Fig. \ref{fig:analogies}.

\subsection{ Fluid Mechanics }

The primary motivation of this work is to study unidirectional (i.e. fully developed, laminar) flow in a straight pipe of arbitrary cross section (Fig. \ref{fig:analogies}(1)). The axial velocity profile $u_z(x,y)$ satisfies Eq. (\ref{eq:flow}) with $k=G/\mu$, where $G$ is (minus) the axial pressure gradient and $\mu$ the viscosity. On the boundary, we assume either no slip, $u=0$ (Poiseuille flow), or, more generally, a prescribed velocity distribution, $u_z=U(x,y)$, for moving walls (Couette flow).  The model can also be extended for hydrodynamic slip, $b\, (\hat{n}\cdot\nabla u_z)=u_z-U$, or stress-free surfaces, $\hat{n}\cdot\nabla u_z = 0$, \change{ such as the upper free surface in simple models of rivers and glaciers, the application for which the c.g.s. viscosity unit ``Poise" was proposed to honor Poiseuille~\cite{deeley1913,sutera1993}. }

The same mathematical problem also describes the circulating flow in a tube of constant vorticity $\omega_z=\nabla\times{\bf u}$  (Fig. \ref{fig:analogies}(2)).  As noted by \citet{greenhill1910}, the stream function $u=\psi(x,y)$, which defines the velocity field, ${\bf u}=-\nabla\times\psi\hat{z}$, satisfies Eq. (\ref{eq:flow}) with $k=\omega_z$.  In this case, the boundary has perfect slip. \change{ Circulating flows in acute-angle corners  \citep{fraenkel1961} have the same scaling as unidirectional corner flows \citep{moffatt1980,collins1976} discussed below. }

Here, we note another analogy of pipe flow with forced gravity currents in porous media \citep{huppert2000}, specifically groundwater flow ~\citep{dunne1980}.  In the \citet{dupuit1863} approximation, the height $h(x,y)$ of groundwater spreading over an flat impermeable rock through a porous soil of hydraulic permeability $\kappa$, fed by a mean precipitation rate $P$, satisfies Eq.~(\ref{eq:flow}) with $u(x,y) = h(x,y)^2$ and $k=2P/\kappa$  (Fig. \ref{fig:analogies}(3)).  Dirichlet boundary conditions, $u=0$, correspond to free drainage out of the soil, e.g. into a river network, and Neumann conditions represent an impermeable boundaries or symmetry lines \citep{cohen2015}.

\subsection{ Solid Mechanics }

Exact solutions of Poisson's equation in two dimensions have an even longer history in solid mechanics \citep{timo_book,musk_book}.  \citet{venant1855} first formulated and solved the general problem of torsion of a prismatic beam (Fig. \ref{fig:analogies}(4)). The beam is clamped on one end and twisted uniformly with angle $\theta z$.  The stress function, $u=\phi(x,y)$, of \citet{prandtl1903} satisfies Eq. (\ref{eq:flow}) with $k=-2G\theta$, where $G$ is the shear modulus, and vanishes on the boundary. \citet{venant1856} also analyzed a prismatic beam under pure bending, clamped at one end, with a transverse force $F$ applied at the other end (Fig. \ref{fig:analogies}(5)). The axial normal stress, $u=\tau_{zz}(x,y)$, satisfies Eq. (\ref{eq:flow}) with $k=F/I(1+\nu)$, where $I$ is the bending moment of inertia and $\nu$ is Poisson's ratio.

\citet{prandtl1903} also introduced the `membrane analogy' for these problems of beam elasticity  (Fig. \ref{fig:analogies}(6)). Equation (\ref{eq:flow}) with $k=\Delta p/\gamma$ now describes the height of a membrane, $u=h(x,y)$, under small elastic deflection by a uniform pressure difference $\Delta p$, and resisted by a constant surface tension $\gamma$.  The same analogy also applies to the interface between two immiscible fluids \citep{griffith1917}, as discussed in one of the earliest papers of G. I. Taylor. In equilibrium, the interface has constant mean curvature, which describes many situations, such as a liquid meniscus~\citep{batchelor_book},  a soap bubble with a pinned contact line~\citep{isenberg_book}, or the Cassie-Baxter state of a textured superhydrophobic surface~\citep{lobaton2007}. For small deflections, $|\nabla h|\ll 1$, the Young-Laplace equation, 
\begin{equation}
-\Delta p=\gamma\  \nabla\cdot \hat{n} = \gamma\  \nabla\cdot \left( \frac{\nabla h}{\sqrt{1+|\nabla h|^2}} \right)
\approx \gamma\  \nabla^2 h
\end{equation}
again reduces to Eq.~(\ref{eq:flow}) with $k=\Delta p/\gamma$. 

\subsection{ Heat and Mass Transfer }

The same problem also naturally arises in transport phenomena, such as heat transfer in a pipe heated uniformly by viscous dissipation or another constant heat source \citep{marco1955,tyagi1966,tyagi1972}. A peculiarity of pipe flow is that the mathematical problem for the fluid velocity is equivalent to that of the temperature profile, $u=T(x,y)$, generated by viscous dissipation (Fig. \ref{fig:analogies}(7)), where $k = g/k_F$ in Eq.~(\ref{eq:flow}) and $k_F$ is the heat conductivity in Fourier's law \citep{shah1978,marco1955,cheng1966,rao1980}.

Another important application is the resistive heating of a straight wire of conductivity $\sigma$ and Fourier coefficient $k_F$  passing a uniform current  density $J$ \citep{deen_book} (Fig. \ref{fig:analogies}(8)). The steady temperature $u=T(x,y)$ satisfies Eq.~(\ref{eq:flow}) with $k=J^2/\sigma k_F$. Boundary conditions could specify isothermal ($T$=constant) or insulating ($\hat{n}\cdot\nabla T=0$) surfaces.  In mass transfer, the same problem could describe the concentration profile $c(x,y)$ of a  chemical of diffusivity $D$ produced by a uniform volumetric reaction rate $R_v$ in a catalyst rod (Fig. \ref{fig:analogies}(9)) \citep{drewry1995}.

\subsection{ Stochastic Processes }

Here, we identify some further analogies of unidirectional flow in stochastic processes, or random walks, which provide the microscopic basis for continuum models of heat and mass transfer.  The mean first passage time $u=\tau(x,y)$ of a Wiener process (i.e. a random walk with infinitesimal, independent, identically distributed displacements with bounded variance) having diffusivity $D$ from a point in a two dimensional domain to its boundary  satisfies Eq.~(\ref{eq:flow}) with $k=D^{-1}$ and an absorbing boundary condition $u=0$ on the target boundary ~\citep{redner_book,vankampen_book} (Fig. \ref{fig:analogies}(10)).   Reflecting boundaries can also be included, with Neumann boundary conditions.  \change{ In finance, first passage processes arise in the pricing of American options or other derivative securities \citep{hull_book}, where a bounded planar domain would describe the range of two underlying assets values where it is not yet profitable to exercise the option.
There are also analogies in polymer physics~\citep{degennes1979}. The same random-walk problem describes the mean length of a polymer, fluctuating inside a tube or disk, from any interior point to a point where it is attached to the wall (Fig. \ref{fig:analogies}(11)).   
}

\subsection{ Diffusion-Controlled Reactions }

\change{
The first passage time (or ``escape time" or ``survival time") describes many phenomena in science and engineering, 
such as the mean reaction time in Smoluchowski's theory of diffusion-controlled homogeneous reactions~\citep{smoluchowski1916,calef1983,redner_book}.  The two-dimensional case  considered here could describe adsorbed reactants on a surface as in heterogeneous catalysis~\citep{drewry1995,dagdug2015} (Fig. \ref{fig:analogies}(12)).  Many applications arise in biology, such as ligand binding on cells or ligand accumulation in cell culture assays ~\citep{lauffenburger_book,berg_book}.  The connection between Poisson's equation and the rate of diffusion-controlled reactions was perhaps first noted by ~\citet{reck1965} in 1965. A general statistical homogenization theory based on the first passage time was formulated by Rubinstein and Torquato for diffusion-controlled reactions\cite{rubinstein1988} and viscous flow~\cite{rubinstein1989}, leading to many mathematical results on reaction-diffusion processes in porous media~\cite{torquato1991,torquato_book}.  
}

\change{ 
In 1990, \citet{torquato1990} unified stochastic processes with viscous flow in arbitrary three-dimensional geometries, a sweeping generalization that has not yet been appreciated for many of the other analogies discussed in this paper.  He established a rigorous link between the mean survival time for diffusion-controlled reactions in a porous medium and the Darcy fluid permeability tensor, ${\bf k}_D$ for Stokes flow in the same porous medium, which relates the mean velocity to the mean pressure gradient, 
\begin{equation}
{\bf u} = - \frac{{\bf k}_D}{\mu} \nabla p.
\end{equation}
In particular, Torquato proved the general inequality,
\begin{equation}
{\bf k}_D \leq \tau \phi_1 D {\bf I} 
\end{equation}
(i.e. $\tau \phi_1 D {\bf I} -{\bf k}_D$ is positive semi-definite), where $\phi_1$ is the porosity and ${\bf I}$ is the identity matrix, and he showed that equality holds, $k_D = \tau \phi_1 D$, in the limit of unidirectional flow in straight parallel pores (as also noted above).  This analogy was exploited by \citet{hunt1995} for the probabilistic computation of Poiseuille flow fields by Monte Carlo simulations of random walks, similar to the algorithm of  \citet{torquato1989} to calculate the effective conductivity of porous media~\cite{kim1990,kim1991}.  The stochastic algorithm enables the efficient approximation of the flow field in pipes with complicated cross sections, including rough, fractal shapes~\cite{hunt1995}.
}

\subsection{ Magnetostatics }

\change{
\citet{heaviside1881} discovered one of the earliest analogies of beam torsion, to the magnetic self-induction of a shielded electrical wire (Fig. \ref{fig:analogies}(13)).  In modern terminology~\cite{jackson_book}, Heaviside expressed the magnetic induction in terms of the vector potential, ${\bf B} = \nabla\times{\bf A}$, chose the Coulomb gauge, $\nabla\cdot{\bf A} = 0$, and considered  uniform current density $J$ in a wire of constant magnetic permeability $\mu_m$, shielded by a perfectly conducting metal sheath.  The magnetic induction circulates around the current in the cross section, and the vector potential plays the role of the stream function in a vortex tube (Fig. \ref{fig:analogies}(2)).   The axial vector potential amplitude $u=A(x,y)$ satisfies Eq. (\ref{eq:flow}) with $k=\mu_m J$ and Dirichlet boundary conditions.  Heaviside exploited this analogy to apply de Saint-Venant's results for torsion to the self-induction of shielded wires of different shapes, but apparently did not recognize the analogy with Poiseuille flow~\cite{heaviside1881}. 
}

\subsection{ Electrostatics }

Of course, the eponymous application of Poisson's equation is in electrostatics, e.g. for a two dimensional cylinder (wire, cavity, etc.) of constant charge density $\rho_e$ and permittivity $\varepsilon$ (Fig. \ref{fig:analogies}(14)). The electrostatic potential $u=\phi(x,y)$ satisfies Eq.~(\ref{eq:flow}) with $k=\rho_e/\varepsilon$ with a prescribed potential  profile on the boundary.  Another variation on this problem with $k=J/\sigma$ is the potential profile of a conducting sheet  of conductivity $\sigma$, cut to a certain shape with bus bars at the edges, sustaining a uniform normal current density $J$ (Fig. \ref{fig:analogies}(15)).  This problem has been used as an analogy to experimentally visualize the profile of elastic deformation of beams in torsion  \citep{waner1953}.

\begin{figure*}
\includegraphics[width=6in]{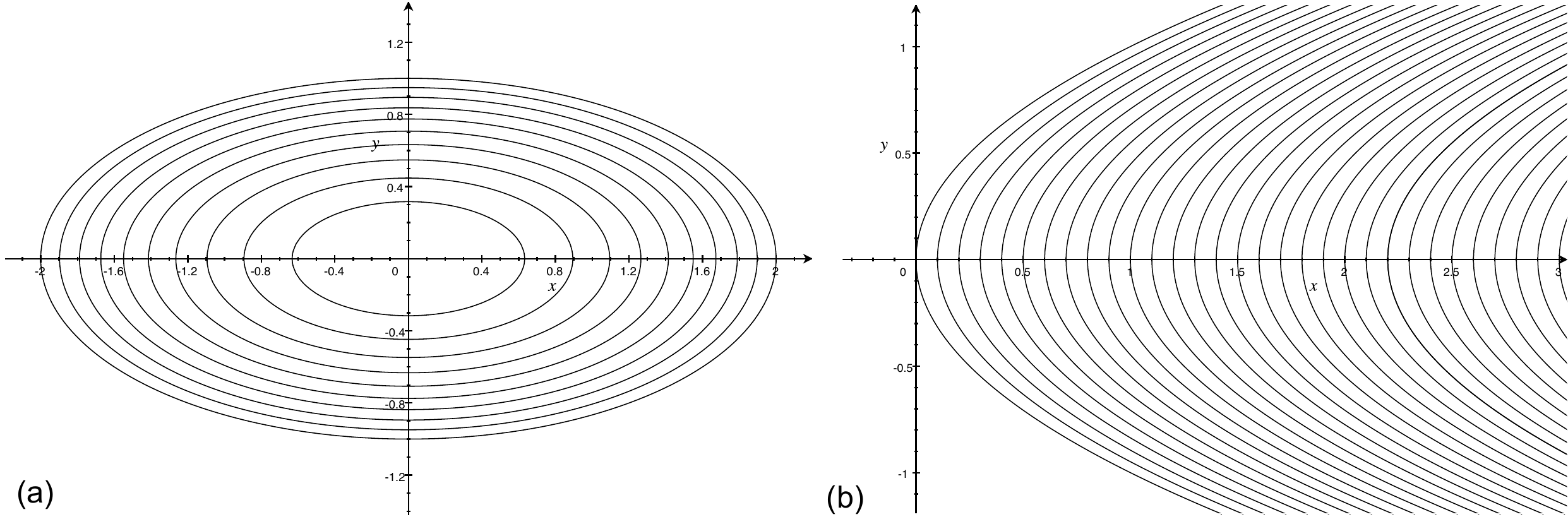}
\caption{  Dimensionless unidirectional flow in (a)   an elliptical pipe ($\tilde{u}=1-(\tilde{x}/2)^2-\tilde{y}^2$)
and (b)  a parabolic groove ($\tilde{u}=\tilde{x}-\tilde{y}^2$).     \label{fig:ell}
}
\end{figure*}

\subsection{ Electrochemistry }

Here, we note that same mathematical problem also describes some problems in electrochemistry.  The steady-state electrostatic potential profile $\phi(x,y)$ satisfies Eq. ~(\ref{eq:flow}) in a planar porous electrode or electrochromic glass~\citep{biesheuvel2011,biesheuvel2012} (Fig. \ref{fig:analogies}(15)), where Dirichlet boundary conditions apply at the current collectors and Neumann boundary conditions at the separator.  This is another variation on the conducting sheet analogy, where again $k=J/\sigma$ but now $\sigma$ is the macroscopic conductivity of the phase with rate limiting transport (electronic or ionic) and $J$ is the Faradaic reaction rate, assumed to be uniform for a perfect analogy.

\subsection{ Electrokinetics }
 
We also note that Poisson's equation arises in three different ways in the theory of electrokinetic phenomena  ~\citep{lyklema_book_vol2,hunter_book}. Besides determining the electrostatic potential profile $\phi(x,y)$ from the charge density $\rho_e$ (i.e. electrostatics, above), it also determines the electro-osmotic flow $u=u_E(x,y)$ in response to an axial electric field with $k=\rho_e E/\mu$ (Fig. \ref{fig:analogies}(16)), as well as the streaming current $u=j_s(x,y)=\rho_e u_p(x,y)$ in response to a pressure-driven flow (Fig. \ref{fig:analogies}(17)).  A perfect analogy with uniform flow requires a constant charge density $\rho_e$, which approximates the diffuse charge profile in a nanochannel with uniform surface charge and thick, overlapping double layers.  \change{  In the absence of flow, the linearized potential profile, satisfying the Debye-Huckel equation for any double-layer thickness, has been analyzed by \citet{duplantier1991} for different two-dimensional shapes, also taking advantage of conformal invariance and connections with random walk theory noted above. } At the end of this article, we shall return to the general situation of non-uniform charge density, $\rho_e(x,y)$, in the full nonlinear problem with electro-osmotic flow.

\section{ Particular solutions for conic section domains }
\label{sec:conic}

In all of these applications, there is a subtle {\it physical} compatibility constraint on the solutions of Eq. (\ref{eq:flow}),
\begin{equation}
uk > 0   \label{eq:sign}
\end{equation}
which requires that $u$ and $k$ have the same sign at each point in the domain.  Here, we will assume $k>0$, so that physical solutions are positive, $u>0$.  In the case of unidirectional flow, the fluid velocity must be directed down the pressure gradient; for resistive heating, a heat source must lead to a rise  in temperature; in electrokinetics, the flow of positive charge is in the direction of the electric field; in soap bubbles, the Laplace pressure is larger on the concave side of the interface; etc.   Since the literature has mostly focused on bounded domains, such as pipe cross sections, this constraint has not been emphasized, but it becomes important when selecting physical solutions in unbounded domains, such as conic sections.


A particular solution of Equation ~(\ref{eq:flow}) is the quadratic form,
\begin{equation}
u(x,y) = \sum_{m=0}^2 \sum_{n=0}^2 A_{mn} x^m y^n    \label{eq:conic}
\end{equation}
which satisfies no slip on a conic section defined by $u=0$, subject to the constraints 
\begin{equation}
2\left(A_{20}+A_{02}\right)=-k \mbox{ \ \ and \ \ } A_{22}=A_{12}=A_{21}=0.   \label{eq:conicA}
\end{equation}  
The allowable domains satisfying Eq. (\ref{eq:sign}) shed light on the physics of unidirectional flow, as well as the other applications sketched in Figure~\ref{fig:analogies}. 


The most important and well known case  is that of an elliptical cross section ($(x/a)^2 + (y/b)^2<1$), introduced by  \citet{venant1855}. The solution is 
\begin{equation}
u(x,y) = K \left\{ 1- \left[ \left(\frac{x}{a}\right)^2 +   \left(\frac{y}{b}\right)^2 \right]  \right\}   \label{eq:ellipse}
\end{equation}
where $K=k/(2(a^{-2} + b^{-2}))$.  An example is shown in Fig. \ref{fig:ell}(a). The limit $a\to\infty$ describes pressure-driven flow between parallel plates ($|y|<b$),
\begin{equation}
u(x,y) = \frac{k}{2} (  b^2 - y^2 )   \label{eq:flat}
\end{equation}
and the case $a=b=R$ corresponds to the Hagen-Poiseuille flow in a circular pipe ($r<R$),
\begin{equation}
u(r) = \frac{k}{4} \left( R^2 - r^2 \right)   \label{eq:hagen}
\end{equation}

Unbounded conic section domains are also included in Eq. ~(\ref{eq:conic}). The simple solution
\begin{equation}
u = \frac{k}{2\kappa}(x-\kappa y^2)
\end{equation}
satisfies no-slip on a parabola ($x=\kappa y^2$). The flow domain ($u>0$) is the region  inside the parabola where the walls provide enough viscous drag to balance the pressure gradient and reach a steady state, as shown in Fig. \ref{fig:ell}(b).

\begin{figure*}
\includegraphics[width=6in]{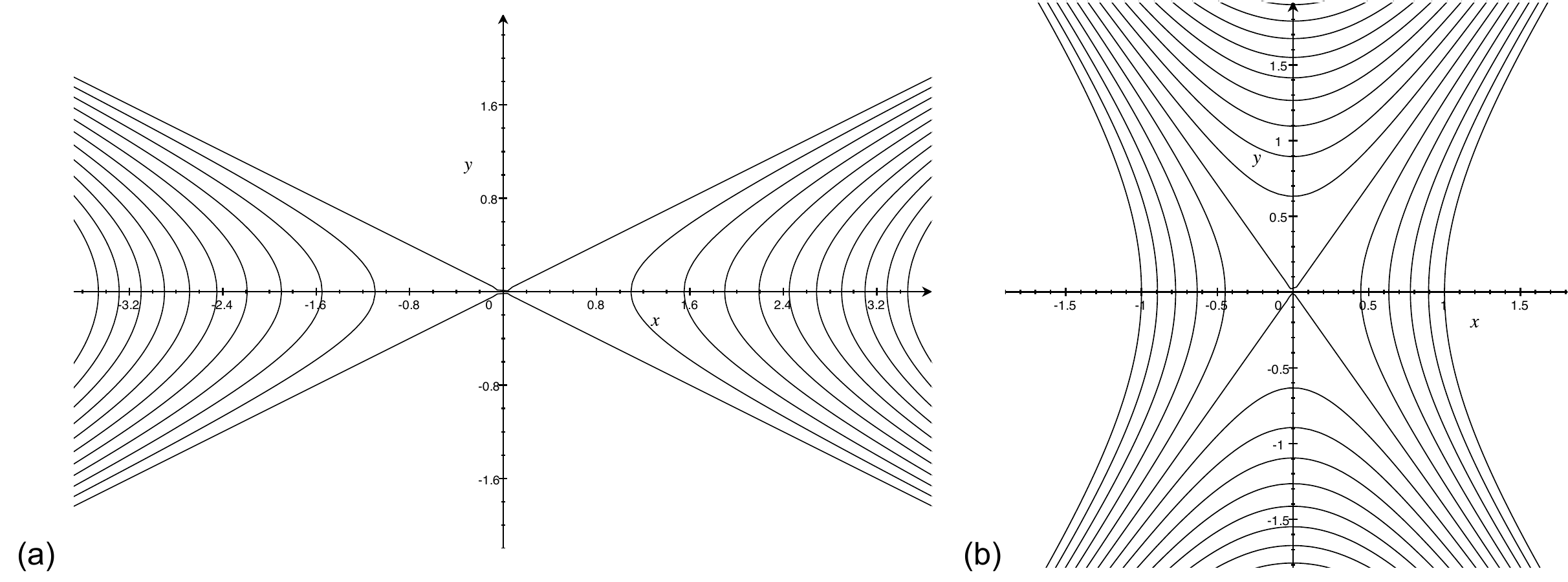}
\caption{  Unidirectional flow in (a) a narrow wedge ($\tilde{u} = (\tilde{x}^2-4\tilde{y}^2)/6$) and (b)  a  hyperbolic constriction ($\tilde{u} = 1 -x^2+y^2/2$). \label{fig:hyp} } 
\end{figure*}


Another simple solution
\begin{equation}
u = A \left[ \left(\frac{x}{a}\right)^2 - \left(\frac{y}{b}\right)^2 \right] + B,  \label{eq:hyper}
\end{equation}
with $A = k/(2(a^{-2} - b^{-2}))$ and $a\neq b$, satisfies no-slip on a hyperbola ($B\neq 0$) or  wedge ($B=0$). 
For $a<b$ and $B=0$ the flow domain lies inside a wedge of acute opening angle $2\tan^{-1}(a/b)<\frac{\pi}{2}$ along the $x$ axis with $|x/a| > |y/b|$, as shown in Fig. \ref{fig:hyp}(a). For $a>b$ and $B=0$, the flow domain is again in an acute-angle wedge, but now along the $y$ axis with $|x/a| < |y/b|$.  Physically, there is no allowable solution for an obtuse angle wedge because it cannot exert enough viscous drag to balance the pressure gradient. \change{ Mathematically, the local similarity solution only exists for an acute corner \citep{moffatt1980}. }


For $B>0$ in Eq. (\ref{eq:hyper}),  the flow domain lies between two branches of a hyperbola, as shown in Fig. \ref{fig:hyp}(b). The velocity profile is a saddle surface, which is a growing parabola along the centerline between the hyperbolic surfaces, and a  decaying parabola along the line of closest approach.   This solution demonstrates the general principle that viscous drag in unidirectional flow is dominated by the narrowest constriction, where the transverse flow profile is approximately parabolic.

The hyperbolic (or wedge) particular solutions can also be written in polar coordinates \change{ \citep{moffatt1980,collins1976,fraenkel1961}},
\begin{equation}
u(r,\theta) = \frac{kr^2}{4}\left( \frac{\cos(2\theta)}{\cos(2\alpha)} - 1 \right) + c.  \label{eq:wpolar}
\end{equation}
The maximum velocity at each radius is along $\theta=0$ and scales as $u \sim x^2$.  The larger scaling exponent than in the case of a parabolic corner, $u \sim x$, reflects the weaker geometrical confinement of the flow in the wedge.   The physical constraint, Eq. (\ref{eq:sign}), implies $\alpha<\frac{\pi}{4}$.  In the case of a wedge ($c=0$), we see again that a physical solution is only possible for acute opening angles, $2\alpha < \frac{\pi}{2}$.

\section{  Equilateral  domains  }

\begin{figure*}
\begin{center}
\includegraphics[width=\linewidth]{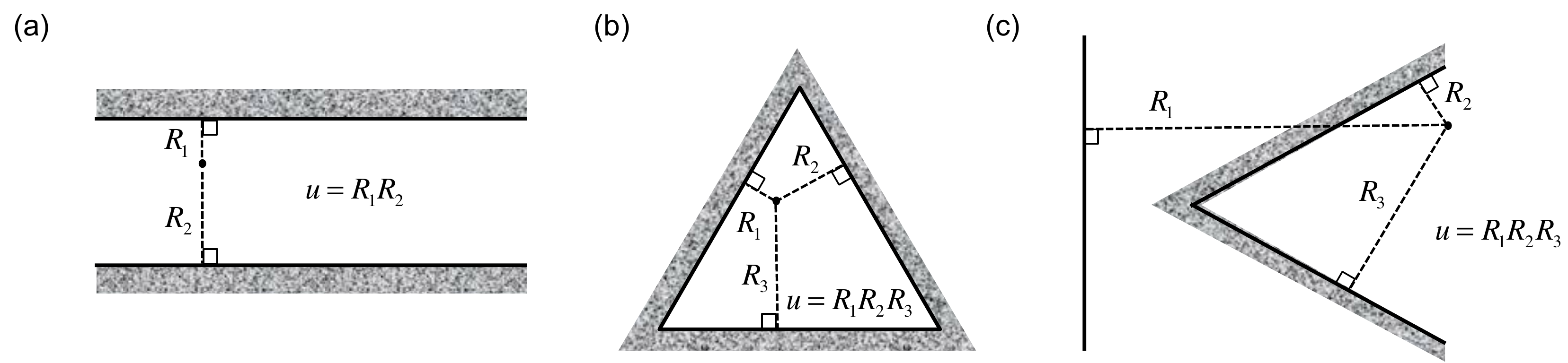}
\caption{ Geometrical construction of special unidirectional flows, obeying Eq. ~(\ref{eq:vec}). 
Famous solutions for parallel plates (a) and equilateral pipes (b) are proportional to the product of the distances to the boundaries. The latter solution also holds for equilateral  wedges (c), where the third boundary lies beyond the apex of the wedge.   \label{fig:equi}  }
\end{center}
\end{figure*}

\subsection{ Geometrical ansatz } 

As beautifully demonstrated by \citet{needham_book}, solutions to Laplace's equation in two dimensions have simple geometrical interpretations, inherited from their analytic complex potentials.  For example, the harmonic function  in the upper half plane with piecewise constant boundary conditions is a weighted sum of angles subtended at the discontinuities, and its harmonic conjugate  is the logarithm of a product of distances to the discontinuities raised to certain powers.  These constructions are related to the Schwartz-Christoffel mapping of the upper half plane to a polygon.  Even more remarkable is Schwarz' geometrical interpretation of Poisson's integral formula, which solves the Dirichlet problem for Laplace's equation in a disk, by averaging the boundary data after circular inversion through an interior point \citep{needham_1994}.

In contrast, solutions of Poisson's equation do not seem to have simple geometrical interpretations -- with a notable exception, the case of an equilateral triangular domain.  Textbooks on fluid mechanics leave it as an exercise to show that  the velocity at an interior point is proportional to the product of the distances to the three sides ~\citep{landau_book,bruus_book}.  Below, we show that the same construction also applies to an unbounded equilateral wedge.  It is  interesting to note that Poiseuille flow between flat plates is also proportional to the product of the distances to the two sides.  

These cases, shown in Fig.~\ref{fig:equi}, suggest that the same geometrical ansatz might also apply to other situations.  
Consider a polygonal domain of $N$ sides having orientation vectors $\{\hat{n}_i\}$ and distances $\{ c_i>0\}$ from the origin. Inspired by the cases $N=2,3$, let us seek  solution of the form,
\begin{equation}
u = a \prod_{i=1}^N \hat{n}_i\cdot(\vec{x}-c_i \hat{n}_i)  \label{eq:vec}
\end{equation}
which is proportional to the product of all the distances from the point $\vec{x}$ to the $N$ sides and thus automatically satisfies the no-slip Dirichlet boundary conditions.    

\subsection{ Equilateral pipe }

\begin{figure*}
\begin{center}
\includegraphics[width=4.2in]{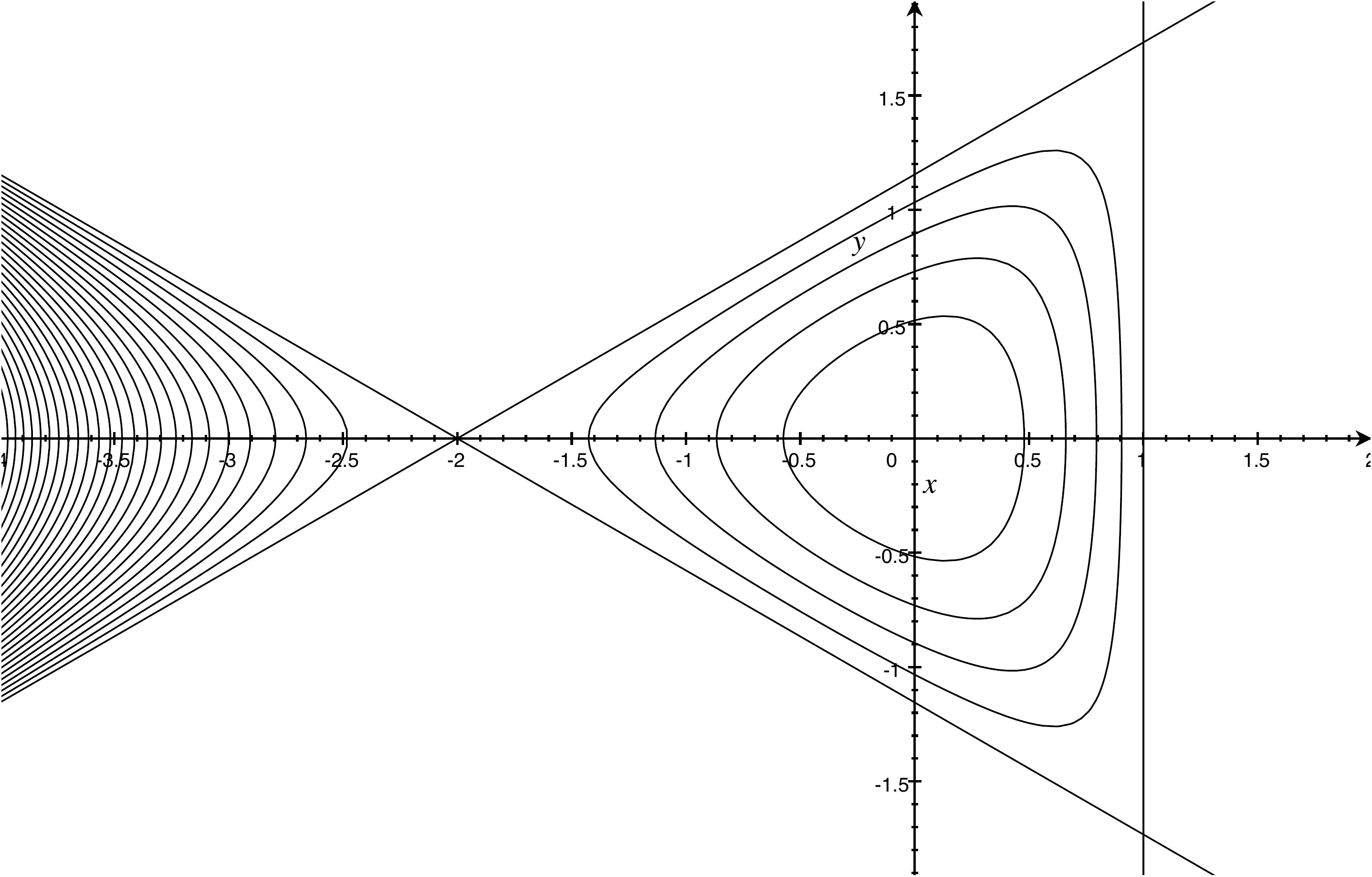}
\caption{  Exact solution for Poiseuille flow in domains with 60$^\circ$ angles between flat no-slip surfaces. Center:  Closed equilateral triangle. Left: Unbounded 60$^\circ$ wedge. The solution in both cases is given by Eq.~(\ref{eq:vec}) for $N=3$ in the domains where $u>0$.  \label{fig:tri}  }
\end{center}
\end{figure*}

First, let us show that for a bounded domain with $N=3$, Eq. (\ref{eq:vec}) only solves Eq. (\ref{eq:flow})  for the case of an equilateral triangle:
\begin{eqnarray*}
u&=&\frac{k}{3c}(x-c)\left[\frac{1}{2}\left(x+\frac{1}{2}\right)-\frac{\sqrt{3}}{2}\left(y-\frac{\sqrt{3}}{2}\right)\right] \\
& & \left[\frac{1}{2}\left(x+\frac{1}{2}\right)+\frac{\sqrt{3}}{2}\left(y+\frac{\sqrt{3}}{2}\right)\right]
\nonumber \\
&=& \frac{k}{12c}(x-c)\left[(2x+4c)^2 - 12y^2\right]    \label{eq:triangle}
\end{eqnarray*}
This well known solution \citep{tao1961,landau_book,bruus_book} is shown in Fig. ~\ref{fig:tri}.

\vskip 12pt

\noindent {\bf Proof:}  
Substituting (\ref{eq:vec}) into (\ref{eq:flow}), we obtain 
\begin{equation}
 \nabla^2 u = a \sum_i \sum_{j\neq i} (\hat{n}_i \cdot \hat{n}_j) \prod_{l \neq i,j} ((\hat{n}_l\cdot\vec{x}) - c_l) = -k\label{eq:lap}
\end{equation}
Since $k$ is constant, all terms involving $\vec{x}$ must vanish. 
For $N=3$:
\[
\nabla^2 u  = 2a ( \vec{b}\cdot \vec{x} + b_0 ) = - k
\]
where 
\begin{eqnarray*}
b_0 &=& c_1(\hat{n}_2\cdot\hat{n}_3) + c_2(\hat{n}_1\cdot\hat{n}_3) +c_3(\hat{n}_1\cdot\hat{n}_2) 
\\
\vec{b}  &=&  \hat{n}_1(\hat{n}_2\cdot\hat{n}_3) + \hat{n}_2(\hat{n}_1\cdot\hat{n}_3) +\hat{n}_3(\hat{n}_1\cdot\hat{n}_2) 
\end{eqnarray*}
Without loss of generality, let $\hat{n}_1 = (1,0)$, $\hat{n}_2 = (\cos\theta_2,\sin\theta_2)$, $\hat{n}_3= (\cos\theta_3,-\sin\theta_3)$. Setting $\vec{b}=\vec{0}$ yields
\begin{eqnarray*}
\sin \theta_2 \cos\theta_2 &=& \sin \theta_3 \cos \theta_3 \\
\sin\theta_2 \sin\theta_3 &=&\cos\theta_2\cos\theta_3+\cos^2\theta_2 +\cos^2\theta_3
\end{eqnarray*}
Requiring $\theta_2,\theta_3\in (0,\pi)$ implies $\theta_2=\theta_3=\theta$ and $\tan^2\theta=3$. and thus $\theta = \pi/3$ or $2\pi/3$. The only possible finite domain is an equilateral triangle, which can be centered on the origin by choosing $\theta = 2\pi/3$ and $c_1=c_2=c_3=c$. Finally, $b_0=-k/a$ implies $a = k/3c$. $\Box$

\subsection{ Equilateral wedge  }

As shown in Fig. \ref{fig:tri}, the same solution (\ref{eq:triangle}) also holds with $u>0$ in an unbounded $60^\circ$ wedge-shaped domain.  The geometric interpretation is the same (product of distances to three surfaces of the equilateral triangle) only now the third surface is hidden behind the apex of the wedge, as shown in Fig. \ref{fig:equi}(c). Since it is the product of three distances, the velocity along the centerline of the equilateral wedge has cubic scaling, $u\propto |x|^3$:
\begin{equation}
u(x,0)=\frac{k}{12c}(x-c)(2x+4c)^2 \sim \frac{kx^3}{3c}, \mbox{  as } x\to-\infty.    \label{eq:eqwedge}
\end{equation}
Below in Eq. (\ref{eq:wedgesol}), we shall explain why this scaling differs from that of the hyperbolic wedge solution, $u \sim x^2$, derived above in Eq. (\ref{eq:wpolar}).

\subsection{  Uniqueness of the equilateral solution }

It is clear that the ansatz (\ref{eq:vec}) cannot hold in general. For example, for $N=4$, the bi-parabolic form $u\propto (a^2-x^2)(b^2-y^2)$ violates Eq.(\ref{eq:flow}).  The ansatz also fails for parabolic Hagen-Poiesuille flow for a circular cross section or any other smoothly curved shape, which can be viewed as a limit of infinitely many infinitesimal sides.  
In fact, no solutions exist of the form (\ref{eq:vec})  for $N>3$.  

\vskip 12pt

\noindent {\bf Proof:}  In order to satisfy Eq. (\ref{eq:lap}) for all $\vec{x}$ in the domain (where $x$ and $y$ vary independently) the coefficients of all terms $a_{m,n}x^my^n$ for $m,n=0,1,..,N-2$ must vanish except for the constant term, which must satisfy $a_{0,0}=-k$. We must therefore satisfy $(N-2)^2$ equations by choosing parameters Eq. (\ref{eq:vec}). There are $N$ coefficients $c_l$ and  $N$ orientation angles for the unit vectors $\hat{n}_i$, which are independent, except for overall rotation and dilation (and rescaling of $u$). Solvability thus requires $2N-2\geq (N-1)^2$, or $N<1+\sqrt{7}<4$.  $\Box$

\vskip 12pt

Therefore, while the ansatz (\ref{eq:vec}) holds in general for $N=2$, it only works for the equilateral geometries for $N>2$, where the unique solution (up to translation, rotation, or dilation) is given by Eq. (\ref{eq:triangle}).  Even isosceles triangular domains have different solutions \citep{sparrow1962}.

\section{ Domains for given unidirectional flows }
\label{sec:implicit} 

\subsection{ General solution of Poisson's equation }

For any geometry and Dirichlet boundary condition, $u=U(x,y)$, the solution to Poisson's equation (\ref{eq:flow}) for unidirectional flow can be expressed as the sum of a pressure-driven Poiseuille flow (particular solution) $u_p$ and a Couette shear flow (homogeneous solution):
\begin{equation}
u(x,y) = u_p(x,y) + u_s(x,y)      \label{eq:ups}
\end{equation}
where we can choose $u_p$ to be any of the exact solutions above, such as Hagen-Poiseille flow (\ref{eq:hagen}), or the general conic section flow (\ref{eq:conic}). The shear flow is a harmonic function, which solves Laplace's equation,
\begin{equation}
\nabla^2 u_s = 0
\end{equation}
with a Dirichlet boundary condition, 
\begin{equation}
u_s(x,y)=U(x,y) - u_p(x,y).   \label{eq:bc}
\end{equation}
which is generally {\it non-constant} even in the case of no slip, $U=0$.
As usual, the harmonic function can be expressed as real (or imaginary) part of an analytic function, the complex potential, $u_s = \mbox{Re} \Phi(z)$, where $z=x+iy$. 

In summary, the general solution of Eq.~(\ref{eq:flow}) has the form
\begin{equation}
u(x,y) = \sum_{m=0}^2 \sum_{n=0}^2 A_{mn} x^m y^n + \mbox{Re } \Phi(x+iy)   \label{eq:gen}
\end{equation}
The basic idea of adding a harmonic function to a quadratic particular solution can be traced to \citet{venant1855}, and general formulae  were perhaps first developed by Muskhelishvili in the 1920s \citep{timo_book,musk_book}.
Using Hagen-Poiseuille flow (\ref{eq:hagen}) as the particular solution, we recover the representation of \citet{tao1961},
\begin{equation}
u(z,\overline{z}) =  -\frac{k}{4} z \overline{z} + \Psi(z) + \overline{\Psi}(\overline{z})   \label{eq:tao}
\end{equation}
where $\overline{z} = x-iy$ is the complex conjugate and $2 \Psi= (k/4)R^2+\Phi$.  This  form of the solution can also be derived by integrating Poisson's equation (\ref{eq:flow}) after a change of variables 
 \citep{tao1961,tyagi1972,rao1980}
\begin{equation}
- \frac{\partial^2 u}{\partial \overline{z} \partial z } = \frac{k}{4}
\end{equation}
related to the complex gradient operator \citep{needham_book,bazant2004}. This solution may seem less general, but it is equivalent to Eq. (\ref{eq:gen}). The bilinear terms in the sum are harmonic, and the quadratic terms have the same form as Eq. (\ref{eq:tao}), since $2x^2 = z\overline{z} + \mbox{Re} z^2$ and $2y^2 = z\overline{z}+\mbox{Re}(iz^2)$.  Nevertheless, as illustrated below, it is easier to start with other particular solutions $u_p$ in Eq. (\ref{eq:gen}) when constructing solutions for domains that do not resemble pipes.

Finding the complex potential $\Phi(z)$ for non-constant Dirichlet boundary conditions in a given domain is possible in theory (Section \ref{sec:geo}), but often challenging in practice.  However, we can easily produce new solutions for unidirectional flow by relaxing constraints on the precise shape of the boundary.  By choosing any harmonic function, $u_s$, we can generate a new exact solution of the form, $u=u_p + u_s$,  from any of the particular solutions above, where the no-slip boundary is defined {\it implicitly} by $u=0$. 

\begin{figure*}
\begin{center}
\vspace{0.1in}
\includegraphics[width=6in]{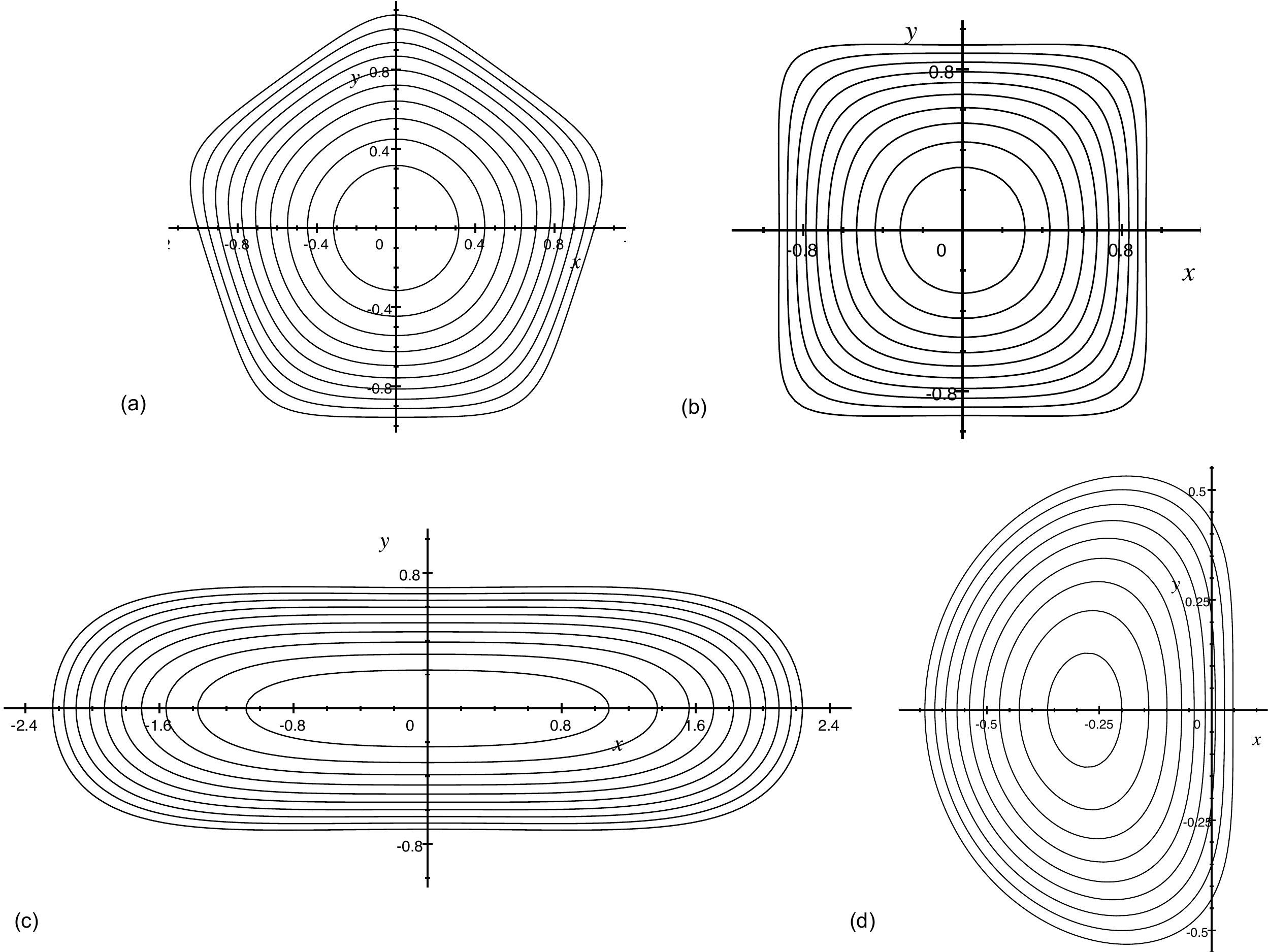}\\
\vspace{0.2in}
\includegraphics[width=4.5in]{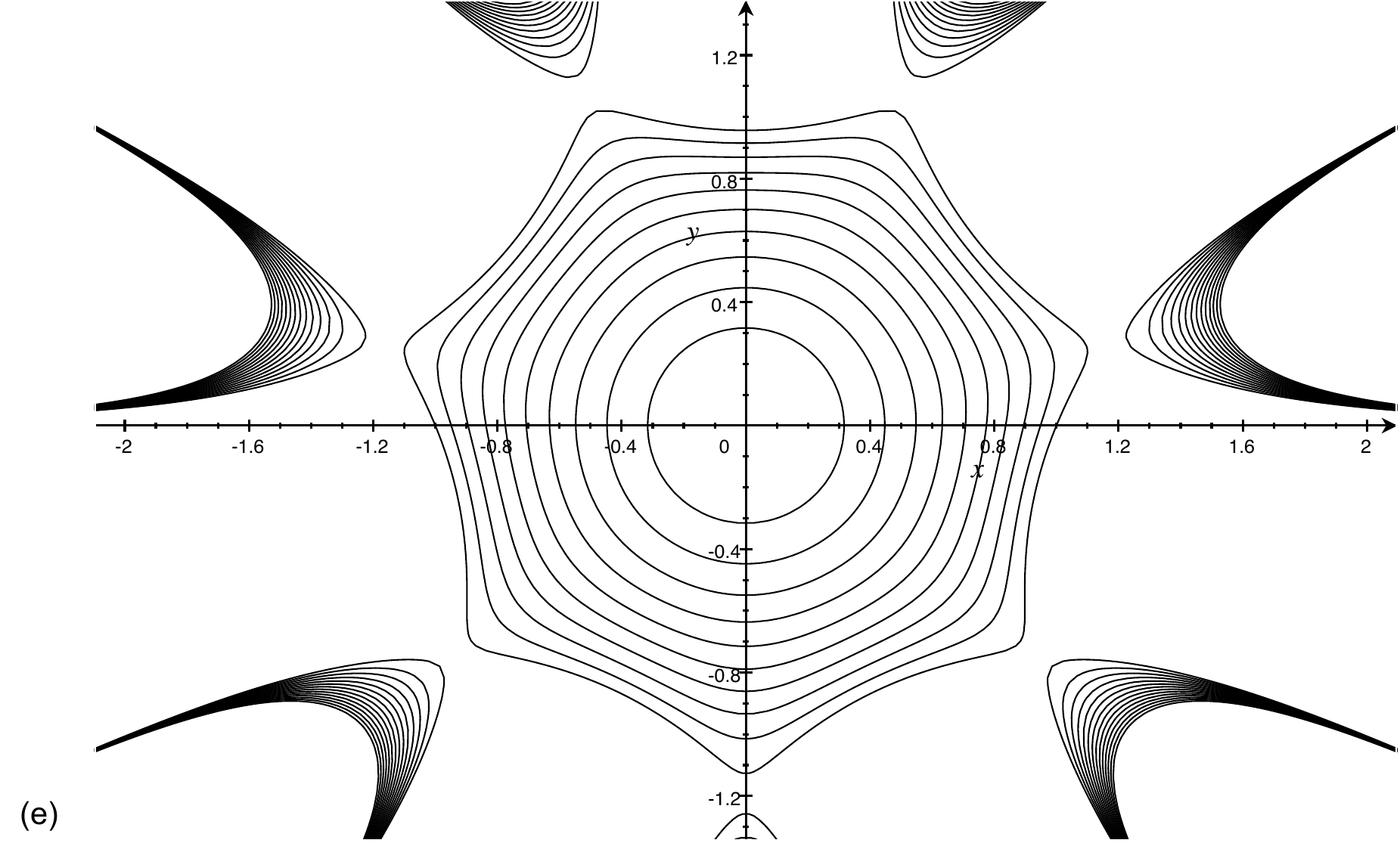}\\
\vspace{0.2in}
\caption{ Exact solutions for unidirectional flow in deformed pipes: (a) a pentagonal pipe and (b) a square pipe given by Eq.~(\ref{eq:rounded}); (c) a rounded slit given by Eq.~(\ref{eq:pertcircle}); (d) an asymmetric ``flattened" pipe, given by Eq. ~(\protect\ref{eq:pole}); and (e) a seven-pointed star pipe, as well as seven smoothed wedge domains, given by Eq.~(\ref{eq:rounded}), with parameters in the main text.
\label{fig:rounded}}
\end{center}
\end{figure*}

\subsection{ Deformed pipes  }

Power-law complex potentials, $\Phi(z) \propto  z^m$, in Eq.~(\ref{eq:tao}) yield Saint-Venant's original solutions for $m-$sided polygonal domains with rounded edges \citep{timo_book}:
\begin{equation}
\tilde{u}= 1 - \tilde{r}^2 + a_m \tilde{r}^m \sin(m\theta)   \label{eq:rounded}
\end{equation}
where we use polar coordinates, $z=r e^{i\theta}$ and define $\tilde{r}=r/R$ and $\tilde{u}=4u/kR^2$.  For example, the no-slip boundary in the case $m=5$ and $a_5=0.11$ resembles a rounded pentagon, and the case $m=4$ and $a_4=-0.21i$ a rounded square, as shown in Fig.~\ref{fig:rounded}(a)-(b).  By varying $a_m$, the  pipe cross section can have more pointed corners, like a star, as shown in Fig.~\ref{fig:rounded}(e) for $m=7$ and $a_7=0.117$.

Interestingly, these exact solutions of \citet{venant1855} resemble  the approximate solutions for slightly different geometries obtained by \citet{mortensen2005} using boundary perturbation methods.  Such approximations have a long history in fluid mechanics, also based on conformal mapping in two dimensions \citep{segel1961a,segel1961b}. These comparisons suggest that, rather than deriving approximate solutions for a given geometry  \citep{bruus_book}, it could be advantageous to derive exact solutions for approximate geometries, guided by the many examples here and in the literature.

A wide variety of other shapes can be created from the general Laurent series, $\Phi(z)=\sum_{n=-\infty}^\infty a_n z^n$, to arbitrary perturbations of the circular geometry:
\begin{equation}
\tilde{u} =  1 - \tilde{r}^2 + \sum_{n=-\infty}^\infty a_n \tilde{r}^n \cos(n\theta)    \label{eq:pertcircle}
\end{equation}
The Laurent coefficients $\{a_n\}$ can in principle be systematically fitted to approximate a given geometry of interest, using the general solution method below.

Let us first consider some simple examples plotted in Fig.~\ref{fig:rounded}.  The choice $a_2=0.93i$ and $a_4=-0.3i$ (with other $a_n=0$) yields a slit-like cross section that closely resembles parallel plates with rounded circular ends, as shown in Fig.~\ref{fig:rounded}(c).   The choice $a_n=-a^{-(n+1)}$ for all $n\geq 0$ distorts the circular pipe from one side,
\begin{equation}
\tilde{u} =  1- \tilde{r^2}-\frac{a-\tilde{r}\cos\theta}{a^2-2a\tilde{r}\cos\theta+\tilde{r}^2}   \label{eq:pole}
\end{equation}
by introducing a simple pole in the complex potential, 
\begin{equation}
\Phi(z)=\frac{1}{z-a}=-\frac{1}{a}\sum_{n=0}^\infty\left(\frac{z}{a}\right)^n.
\end{equation}
If the pole lies just outside the circular domain of unperturbed Hagen-Poiseuille flow, the pipe  appears to be ``flattened" against a planar surface, as shown in Fig.~\ref{fig:rounded}(c) for $a=1.05$.

\change{ The Laurent series (\ref{eq:pertcircle}) corresponds to an implicit parameterization of the boundary ($u=0$) in polar coordinates. \citet{wang1995} showed that exact solutions can also be constructed for a class of symmetric bi-polar shapes, albeit in the context of torsion of a compound bar.
}

\begin{figure*}
\begin{center}
\vspace{0.1in}
\includegraphics[width=6in]{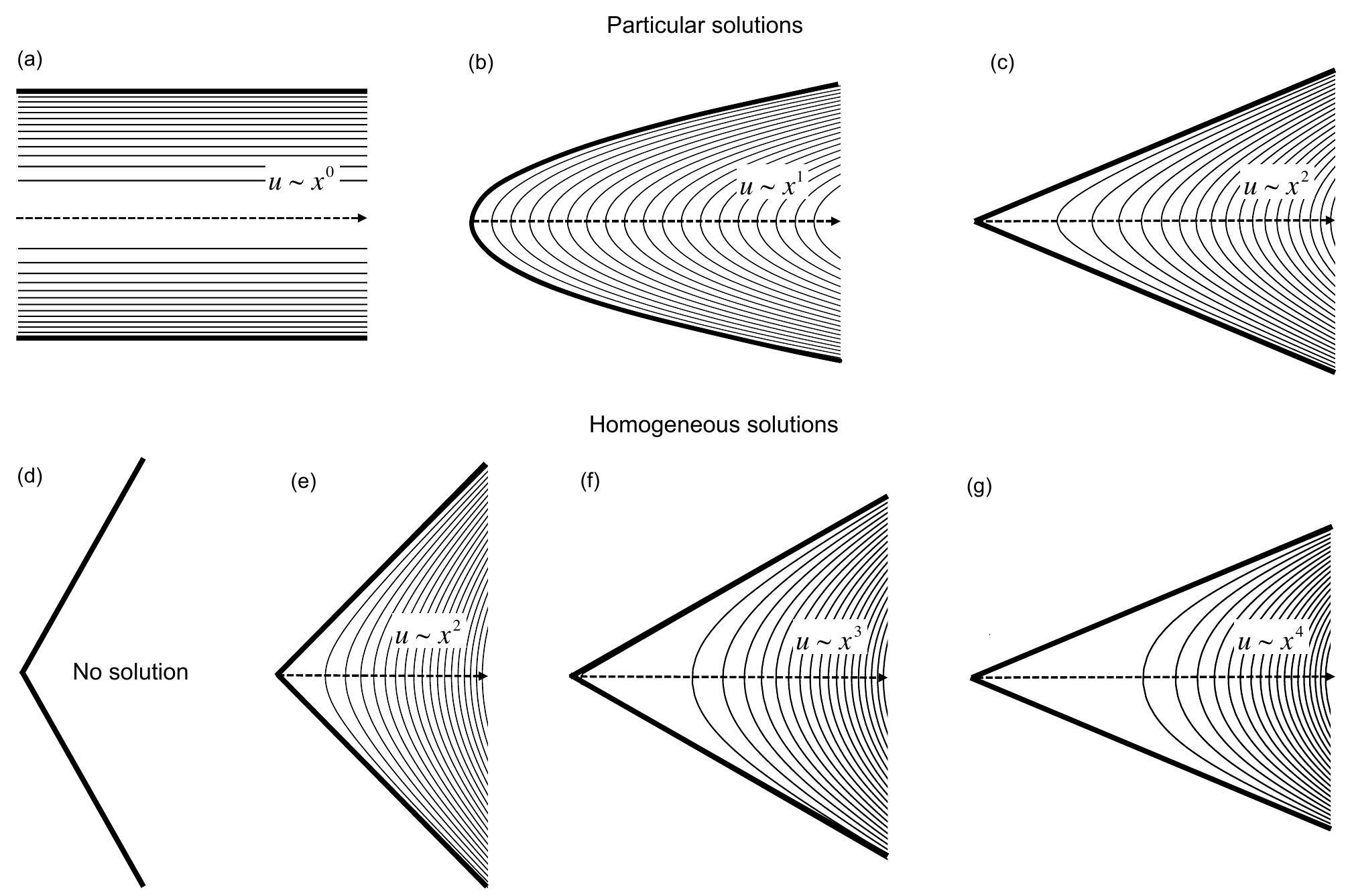}  
\vspace{0.05in}
\caption{ Geometry dependence of the centerline velocity scaling, $u \sim x^p$, near a corner in unidirectional flow.  Particular solutions grow faster (larger $p$) with decreasing confinement, as illustrated by (a) parallel plates with $p=0$, (b) parabolic corners with $p=1$, and (c) acute-angle wedges with $p=2$. No solutions exist for (d) obtuse angle wedges.  For acute-angle wedges, homogeneous solutions instead grow faster with increasing confinement, as shown for opening angles (e) $\pi/2$ with $p=2$, (f) $\pi/3$ with $p=3$ and (f) $\pi/4$ with $p=4$.
\label{fig:corner}}
\end{center}
\end{figure*}

\subsection{  Corners }

Fig. \ref{fig:rounded}(e) also illustrates how these solutions for bounded star-like domains also describe (multiple) unbounded wedge-like domains with smoothed corners, farther from the origin.  The asymptotic scaling of the velocity along the center of each wedge is $u\sim r^m$, where $2\alpha=\pi/m$ is the opening angle of the wedge.  For $m=3$, we recover the asymptotic scaling, $u\sim r^3$,  derived above for the equilateral wedge. The constraint for a positive solution, $m>2$, implies that the wedge angle must be acute, $2\alpha < \pi/2$, as noted above.

The different asymptotic scalings for the hyperbolic and equilateral wedge solutions above can be reconciled by constructing the general solution for a wedge,
\begin{equation}
u(r,\theta) = \frac{k\,r^2}{4}\left( \frac{\cos(2\theta)}{\cos(2\alpha)} - 1 \right) + K r^{\pi/2\alpha} \cos\left(\frac{\pi\theta}{2\alpha}\right)   \label{eq:wedgesol} 
\end{equation}
where the physical constraint, Eq. (\ref{eq:sign}), requires an acute opening angle, $2\alpha < \pi/2$.  
The first term is the hyperbolic particular solution, Eq. (\ref{eq:wpolar}), scaling as $u\sim x^2$ along the centerline. The second term is the homogeneous solution, which always dominates the asymptotic scaling, $u \sim x^{\pi/2\alpha}$ as $x \to \infty$, as a result of the physical constraint.   Since the homogenous solution must be included to satisfy any additional boundary conditions, this scaling dominates in bounded wedge domains, such as the isosceles triangle, where it appears as the leading term in an eigenfunction series solution~\cite{sparrow1962}.   The equilateral wedge solution, Eq. (\ref{eq:eqwedge}), is a special case of Eq. (\ref{eq:wedgesol}), where $u \sim  x^3$ for $2\alpha=\pi/3$.  

The asymptotic scalings of corner flows are summarized in Fig. ~\ref{fig:corner}.  Particular solutions grow faster with decreasing confinement, as $u \sim x^p$ with $p=0$ for parallel plates, $p=1$ for a parabolic corner, and $p=2$ for all wedges with acute opening angles ($2\alpha < \frac{\pi}{2}$).  In contrast, homogeneous solutions grow faster with increasing confinement, as for wedges of decreasing $\alpha$, where $u \sim x^{\pi/2\alpha}$. For wedges with obtuse opening angles, neither particular nor homogeneous solutions exist.

\change{ \citet{moffatt1980} were the first to recognize the significance of right angle corners for Poiseuille flows in bounded ducts, related to the discussion above.  For acute angles, the dominant flow near the corner is a locally determined similarity solution, Eq. (\ref{eq:wpolar}), with sub-dominant terms affected by the global geometry.  For obtuse angles, a local similarity solution does not exist, and the dominant flow is a non-universal eigenfunction expansion, determined globally by the shape of the duct.  Physically, an obtuse angle corner does not exert enough drag on the fluid to control the flow profile.
}

\begin{figure*}
\begin{center}
\vspace{0.1in}
\includegraphics[width=6in]{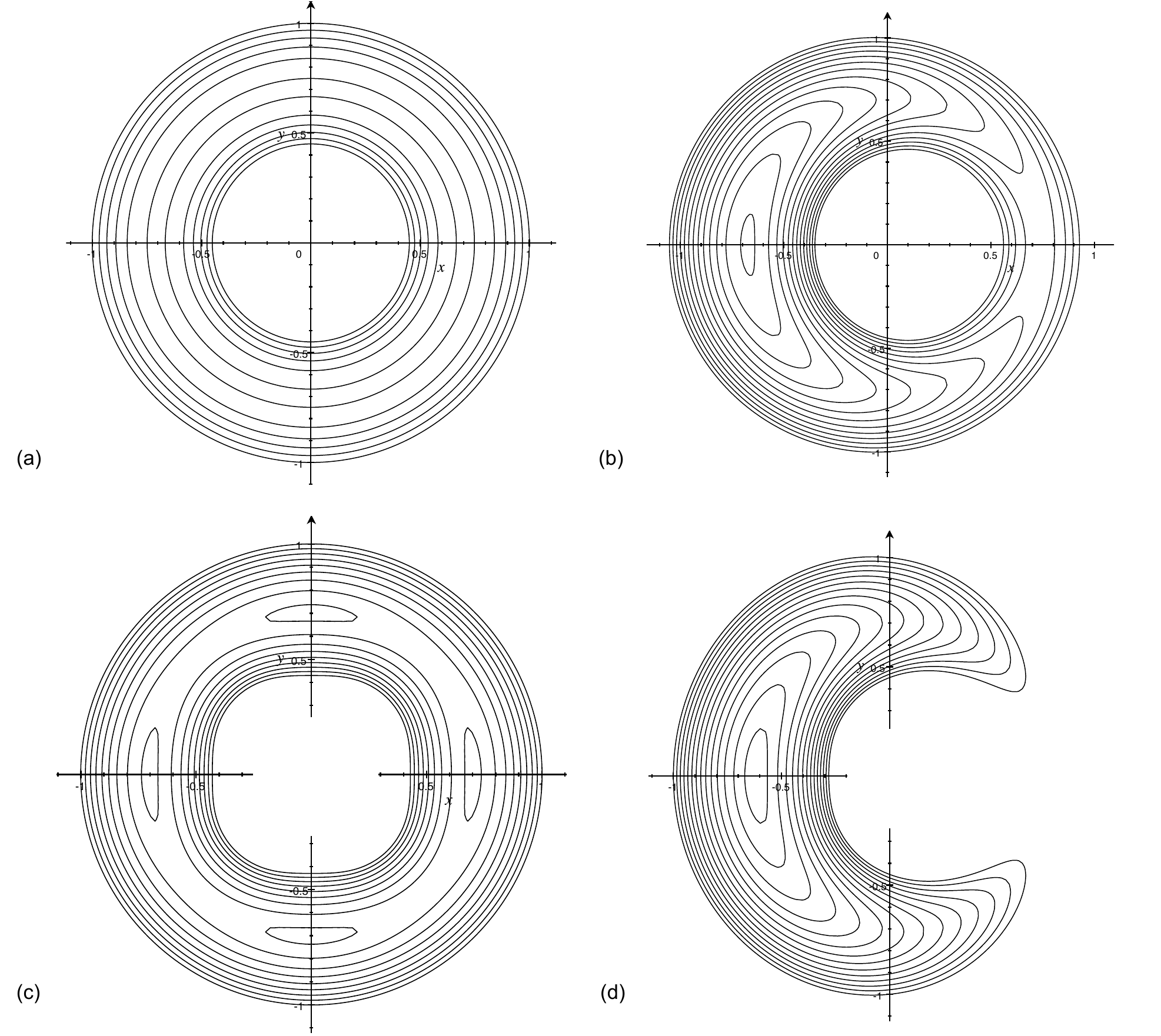}  
\vspace{0.05in}
\caption{ Exact solutions for (a) coaxial circular pipes, Eq.~  (\ref{eq:coax}), as well as (b) off-center near-circular pipes, (c) a rounded square inside a circle, and (d) a single C-shaped single pipe, given by various cases of Eq.~(\ref{eq:acoax}) described in the text.
\label{fig:coax}}
\end{center}
\end{figure*}

\subsection{ Deformed coaxial cylinders }

The domain  for the general solution (\ref{eq:gen}) need not be simply connected. For example by adding shear flow between moving coaxial cylinders, $\Phi(z)\propto \log z$, to Hagen-Poiseuille flow in a circular pipe, we obtain the general solution for unidirectional flow between no-slip coaxial circular cylinders, 
\begin{equation}
\tilde{u} = 1  - \tilde{r}^2 - \beta \ln \tilde{r}   \label{eq:coax}
\end{equation}
shown in Fig.~\ref{fig:coax}(a) for $\beta=1$. Once again, we can add a Laurent series $\Phi(z)=\sum_n a_n z^n$ to represent general perturbations of the coaxial geometry,
\begin{equation}
\tilde{u} = 1 - a_x \tilde{x}^2 - a_y\tilde{y}^2 - \beta \ln \tilde{r} + \sum_{n=-\infty}^\infty a_n \tilde{r}^n \cos(n\theta).   \label{eq:acoax}
\end{equation}
The case $a_x=a_y=\beta=1$, $a_{-1}=0.06$ (with $a_n=0$ for $n\neq -1$), shown in Fig.~\ref{fig:coax}(b), approximates the important case of a misaligned inner cylinder, causing increased hydraulic resistance on the right and faster flow on the left \citep{piercy1933}. The inner pipe can also be distorted into symmetric shapes of $n$-fold symmetry via higher-order poles in the Laurent series. The case $a_x=a_y=\beta=1$ and $a_{-4}=0.001$, shown in Fig. ~\ref{fig:coax}(c), corresponds to a rounded square inner pipe with four velocity maxima along the widest sections centered on the faces. By breaking symmetry in the particular solution, $a_x\neq a_y$, the outer pipe can be more strongly distorted or connected to the inner pipe.  The case $a_x=1.1$, $a_y=\beta=1$, and $a_{-1}=0.1$, shown in Fig.~\ref{fig:coax}(d), creates a single pipe with an asymmetric C-shaped cross section.

\subsection{ Deformed plates }

\begin{figure*}
\begin{center}
\vspace{0.1in}
\includegraphics[width=\linewidth]{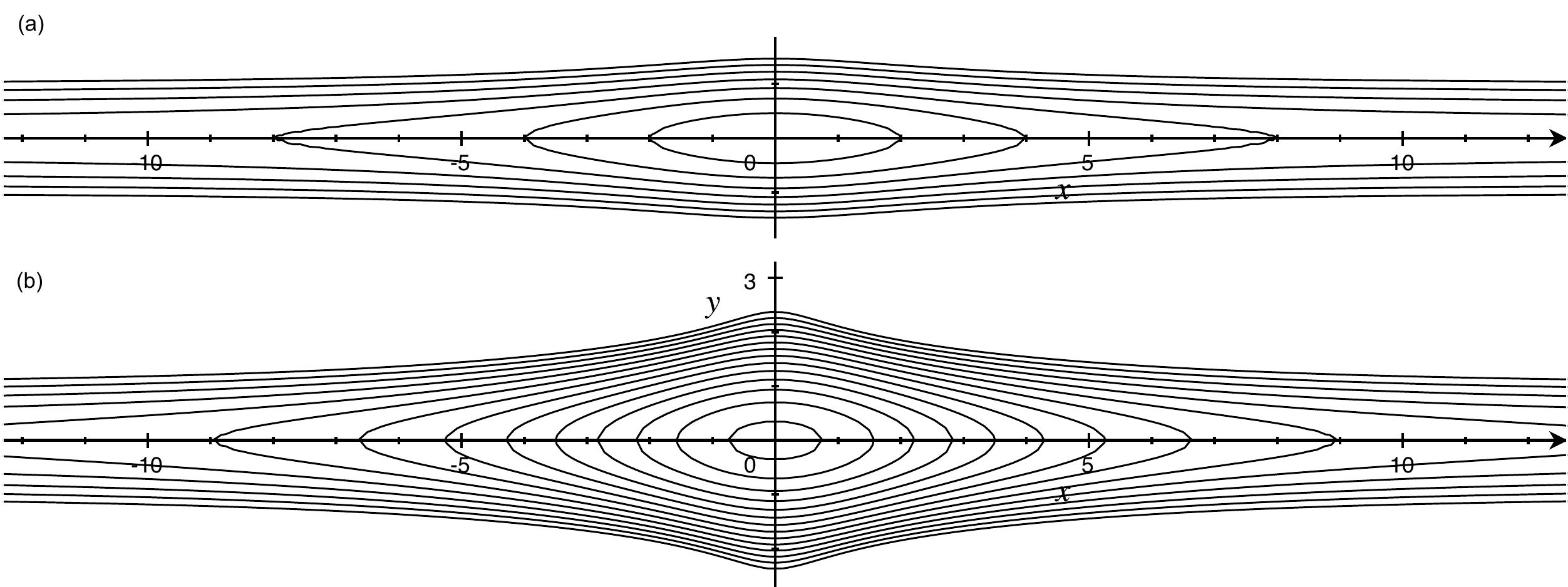}  
\vspace{0.1in}
\caption{ Exact solutions for flow between deformed parallel  plates with a central `bulge', given by Eq. (\ref{eq:bulge}) with $a=4$ and (a) $\gamma=2$, (b) $\gamma=4$, (c) $\gamma=6$.
\label{fig:bulge}}
\end{center}
\end{figure*}

Another class of solutions of the general form (\ref{eq:gen}) corresponds to geometrical perturbations of Poiseuille flow between flat plates,
\begin{equation}
\tilde{u}(\tilde{x},\tilde{y}) = 1 - \tilde{y}^2 +  \mbox{Re } \Phi(\tilde{x}+i\tilde{y})
 \label{eq:wavy}
\end{equation}
where first we consider placing symmetric pairs of simple poles just outside the plates using the function,
\begin{equation}
F(\tilde{z}) = \frac{1}{z-ia} - \frac{1}{z+ia} = \frac{2ia}{z^2+a^2}   \label{eq:F}
\end{equation}
With an imaginary coefficient, $\Phi(\tilde{z})=- i \gamma F(\tilde{z})$, the solution
\begin{equation}
\tilde{u}(\tilde{x},\tilde{y}) = 1 - \tilde{y}^2 + \frac{2a\gamma(\tilde{x}^2-\tilde{y}^2+a^2)}
{(\tilde{x}^2+\tilde{y}^2)^2 + 2a^2(\tilde{x}^2-\tilde{y}^2)+a^4}     \label{eq:bulge}
\end{equation}
describes a symmetric `bulge' in the plates, as shown in Fig. ~\ref{fig:bulge}.   The idea of warping a plate-like geometry by placing poles away from the axis has also been used to construct exact solutions to the Navier-Stokes equations having steady vortex structures \citep{bazant2005}.

\begin{figure*}
\begin{center}
\vspace{0.1in}
\includegraphics[width=\linewidth]{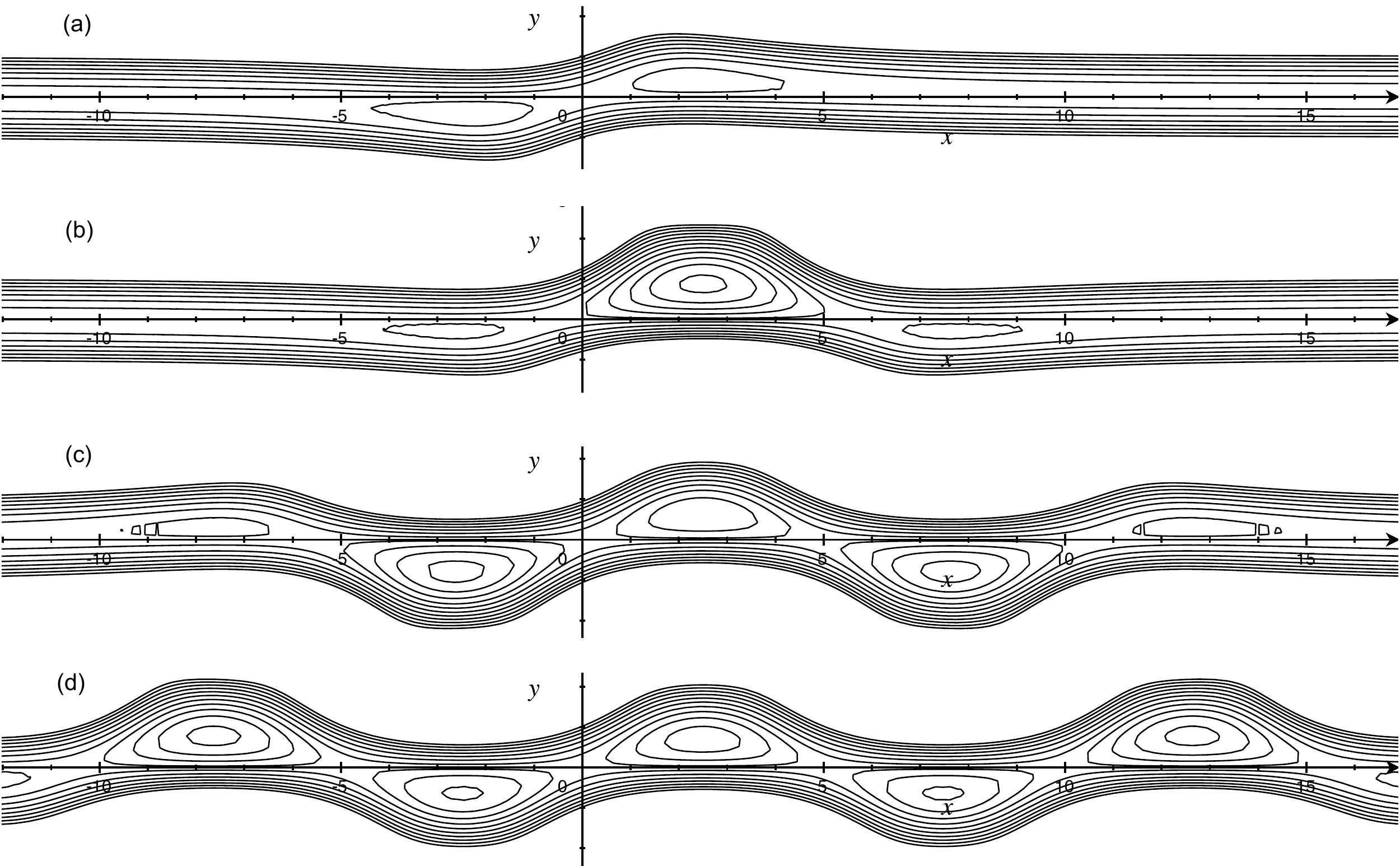}  
\vspace{0.1in}
\caption{ Exact solutions for flow between deformed parallel  plates with a sequence of `twists', given by Eq. (\ref{eq:twist}) with $a=4$, $b=4$ and $\gamma=10$  for (a) one, (b) two, (c) four, and (d) six terms in Eq. (\ref{eq:snake}) .
\label{fig:twist}}
\end{center}
\end{figure*}

Alternatively, with a real coefficient, $\Phi(\tilde{z})=\gamma F(\tilde{z})$, the solution
\begin{equation}
\tilde{u}(\tilde{x},\tilde{y}) = 1 - \tilde{y}^2 + \frac{4a\gamma\tilde{x}\tilde{y}}
{(\tilde{x}^2+\tilde{y}^2)^2 + 2a^2(\tilde{x}^2-\tilde{y}^2)+a^4}     \label{eq:twist}
\end{equation}
describes a `twist' in the plates, as shown in Fig. ~\ref{fig:twist}(a) for $a=4$ and $\gamma=10$.   The plates can then be deformed into a snake-like pattern by adding alternating twists at periodic locations,
\begin{eqnarray}
\Phi(\tilde{z}) &=& \gamma( F(\tilde{z}) - F(\tilde{z}-b) - F(\tilde{z}+b) + F(\tilde{z}-2b)  \nonumber     \\
& &  + F(\tilde{z}+2b) - F(\tilde{z}-3b) +\ldots)  \label{eq:snake}
\end{eqnarray}
as shown in Fig. ~\ref{fig:twist} for $a=4$, $b=5$, and $\gamma=10$.

\begin{figure*}
\begin{center}
\vspace{0.1in}
\includegraphics[width=\linewidth]{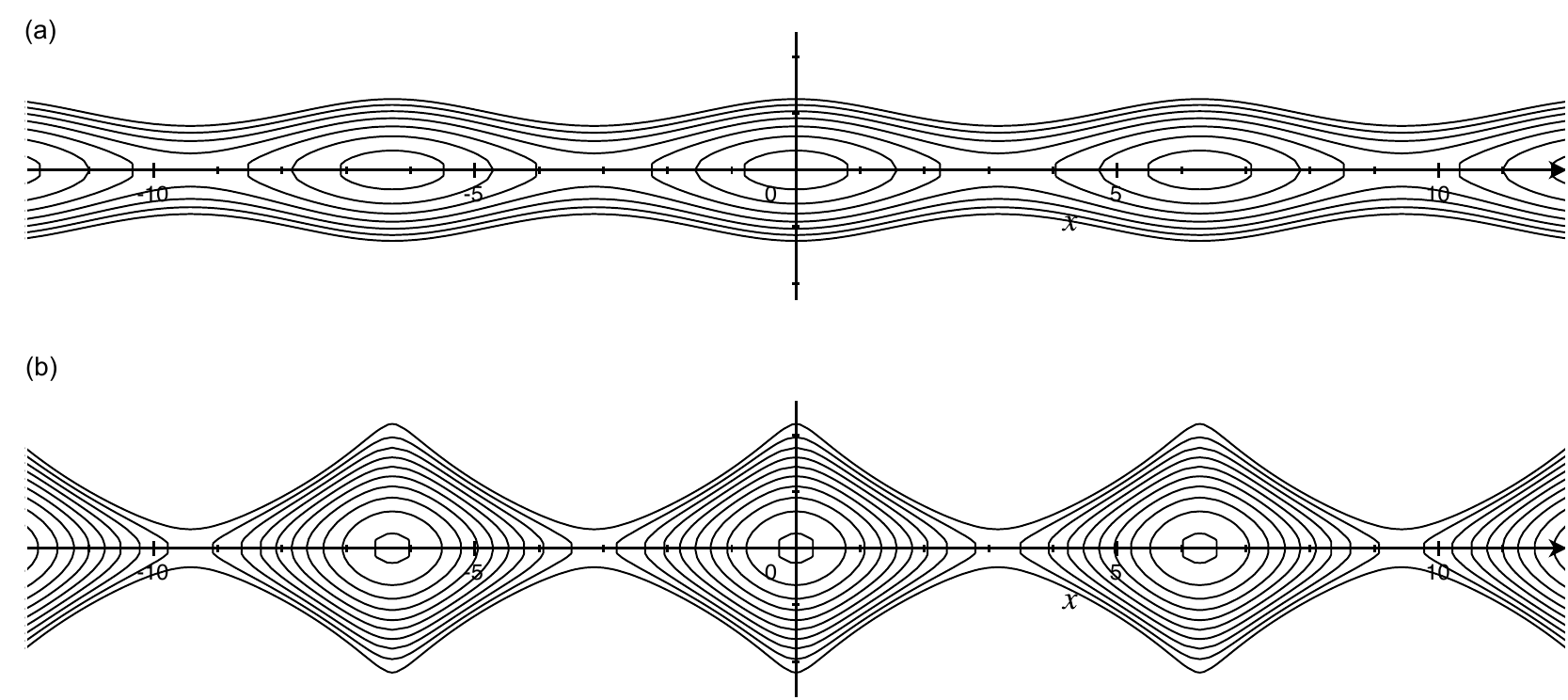}  
\caption{ Exact solutions for flows between periodically grooved plates, given by (\ref{eq:grooved}) with (a) $\sigma=0.3$ and (b) $\sigma=0.84$.
\label{fig:grooved}}
\end{center}
\end{figure*}

Periodic deformations of parallel plates can be achieved by adding complex sinusoidal perturbations.  For example, the exact solution
\begin{equation}
\tilde{u}(\tilde{x},\tilde{y}) = 1 - \tilde{y}^2 +  \mbox{Re }\sigma \cosh(i\tilde{z}) = 1 - \tilde{y}^2 +  \sigma \cosh(\tilde{y})\cos(\tilde{x})  
\label{eq:grooved}
\end{equation}
is shown in Fig.~\ref{fig:grooved}. The case $\sigma=0.3$ in (a) describes smooth grooves along the plates, while the case $\sigma=0.84$ in (b) describes a sawtooth surface with diamond-like ribs.

\begin{figure*}
\begin{center}
\vspace{0.1in}
\includegraphics[width=\linewidth]{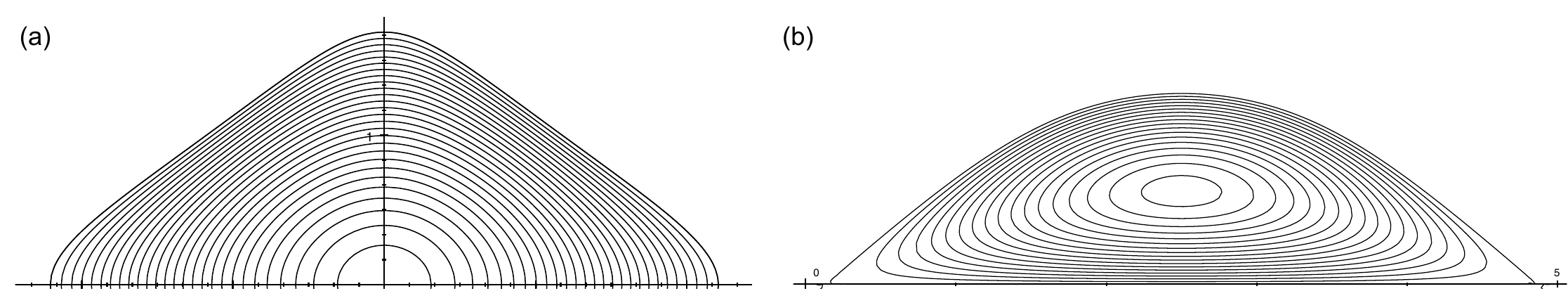} 
\vspace{0.1in}
\caption{ Exact solutions for flows in flat-bottom microchannels, obtained by adding constants to the solutions in (a) Fig.~\ref{fig:twist}(c) and (b) Fig.~\ref{fig:grooved}(b).
\label{fig:shifted}}
\end{center}
\end{figure*}

\subsection{ Flat-bottom microchannels, troughs, and glaciers }

A useful observation is that every isovelocity contour ($u=$constant) in our solutions above could also serve as the no-slip boundary of the domain. This follows from the simple fact that the solution to Eq. (\ref{eq:flow}) is invariant to shifting by a constant, which is a solution to the homogeneous  equation. 

Some examples are shown in Fig.~\ref{fig:shifted} with relevance for microfluidic devices \citep{bruus_book,tabeling_book,kirby_book}. By subtracting 1 from the solution in Fig.~\ref{fig:twist}(c), we obtain the solution in Fig.~\ref{fig:shifted}(a) for a microchannel with a flat floor and a curved ceiling, similar to what can be made by various soft lithography methods. This solution also resembles  the case of a parabolic microchannel analyzed by \citet{bruus_book} using perturbation methods.

As noted above, same mathematical methods could be applied not only to Poiseuille flows in channels with no-slip walls, but also to unidirectional Couette flows with moving walls, or to mixed Couette-Poiseuille flows with both slipping and non-slipping surfaces.
In order to illustrate the latter case, we subtract 0.5 from the solution in Fig.~\ref{fig:grooved}(b) to obtain the flow in Fig.~\ref{fig:shifted}(b) for a triangular channel with a no-slip ceiling and a perfect-slip (stress-free) floor, e.g. due to a superhydrophobic coating. Turned upside down, this is also an example of drainage flow with a free surface in a trough \citep{shah1978}, which could also describe an advancing glacier in a wedge-shaped valley of constant cross section~\cite{deeley1913}.  The examples in Fig. \ref{fig:grooved}, if restricted to the lower half plane, could likewise describe continental ice cap with a flat upper surface, slowly flowing over periodic mountain ranges.

\begin{figure*}
\begin{center}
\vspace{0.1in}
\includegraphics[width=\linewidth]{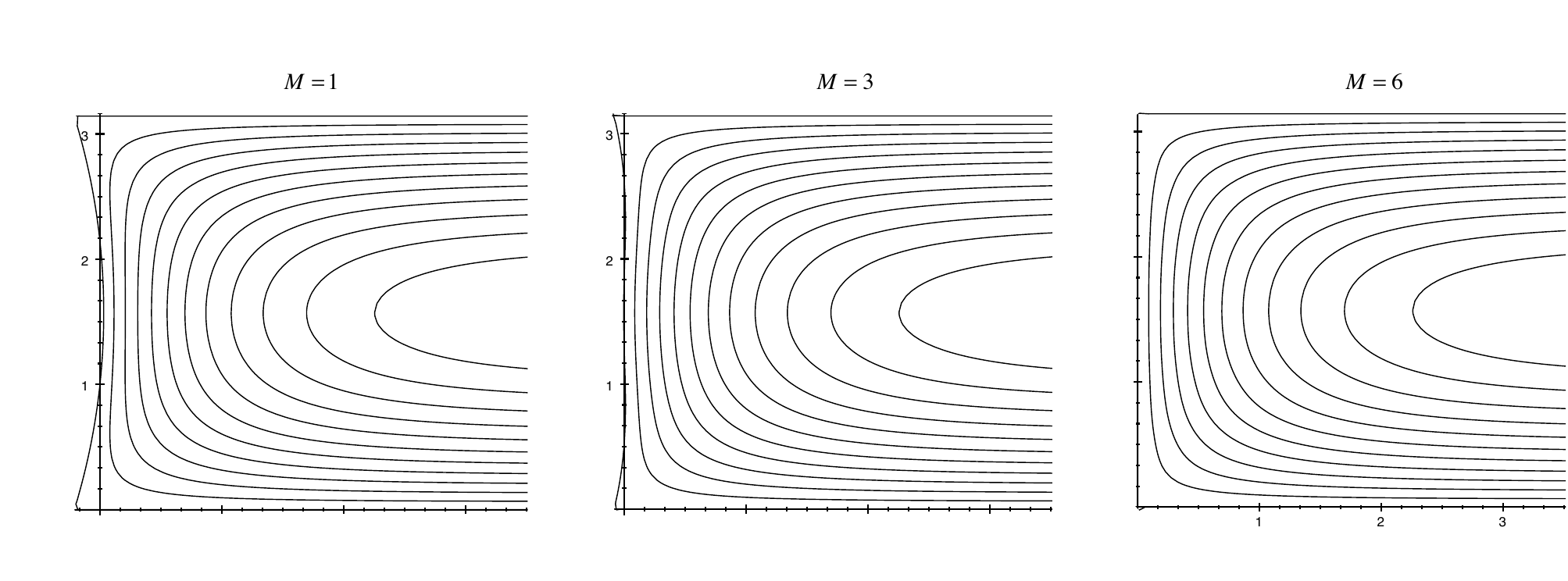}  
\caption{ Exact solutions for nearly rectangular corners with $M$ terms in Eq. (\ref{eq:rectcorn}). 
\label{fig:rectcorn}}
\end{center}
\end{figure*}

\subsection{ Nearly  rectangular domains }

Next we consider rectangular domains, which have well-known series solutions \citep{bruus_book}, such as a rectangular corner (semi-infinite strip), $0 < \tilde{y} < i\pi$, $x>0$, with no slip, $U=0$.  In this case, the solution can be represented as a Fourier series:
\begin{equation}
\tilde{u} = \frac{4}{\pi}  \sum_{n >0\ odd}^\infty \frac{(1-e^{-n\tilde{x}}) \sin(n\tilde{y})}{n^3} 
\end{equation}
where $\tilde{u}=kL^2$, $\tilde{x}=x/L$, $\tilde{y}=y/L$ for a given length scale $L$.
This series cannot be differentiated term by term, e.g. to calculate the viscous stress along the end, but it can be made uniformly convergent (and thus differentiable) by subtracting the particular solution for parallel plates,
\begin{eqnarray}
\tilde{u}_p &=& \frac{\tilde{y} (\pi-\tilde{y})}{2}  = \tilde{u}-\tilde{u}_s  \label{eq:uplate} \\
\tilde{u}_s &=&- \frac{4}{\pi}  \sum_{n >0\ odd}^\infty \frac{e^{-n\tilde{x}} \sin(n\tilde{y}) }{n^3}= \frac{4}{\pi}\  \mbox{Im } \Phi(\tilde{z})  \label{eq:corn}
\end{eqnarray}
In this case, we can sum the series to obtain simple formulae for higher derivatives of the complex potential:
\begin{eqnarray}
\Phi(\tilde{z}) &=&  \sum_{n >0\ odd}^\infty \frac{e^{-n\tilde{z}}}{n^3} \\
\Phi^{\prime\prime\prime}(\tilde{z})
&=& -\sum_{n >0\ odd}^\infty e^{-n\tilde{z}}  \nonumber \\
&=&- \frac{e^{-\tilde{z}}}{1- e^{-2\tilde{z}}}   =- \frac{1}{2\sinh\tilde{z}} \ \  \ \ (\mbox{Re}\  \tilde{z}>0)   \nonumber \\
\Phi^{\prime\prime}(\tilde{z}) &=& -\frac{1}{2}\ln\tanh \frac{\tilde{z}}{2}
\end{eqnarray}
but it does not seem possible to integrate two more times to obtain a closed-form solution. Below we shall encounter similar problems in evaluating the Schwarz integral formula for $\Phi(\tilde{z})$.

\begin{figure*}
\begin{center}
\vspace{0.1in}
\includegraphics[width=\linewidth]{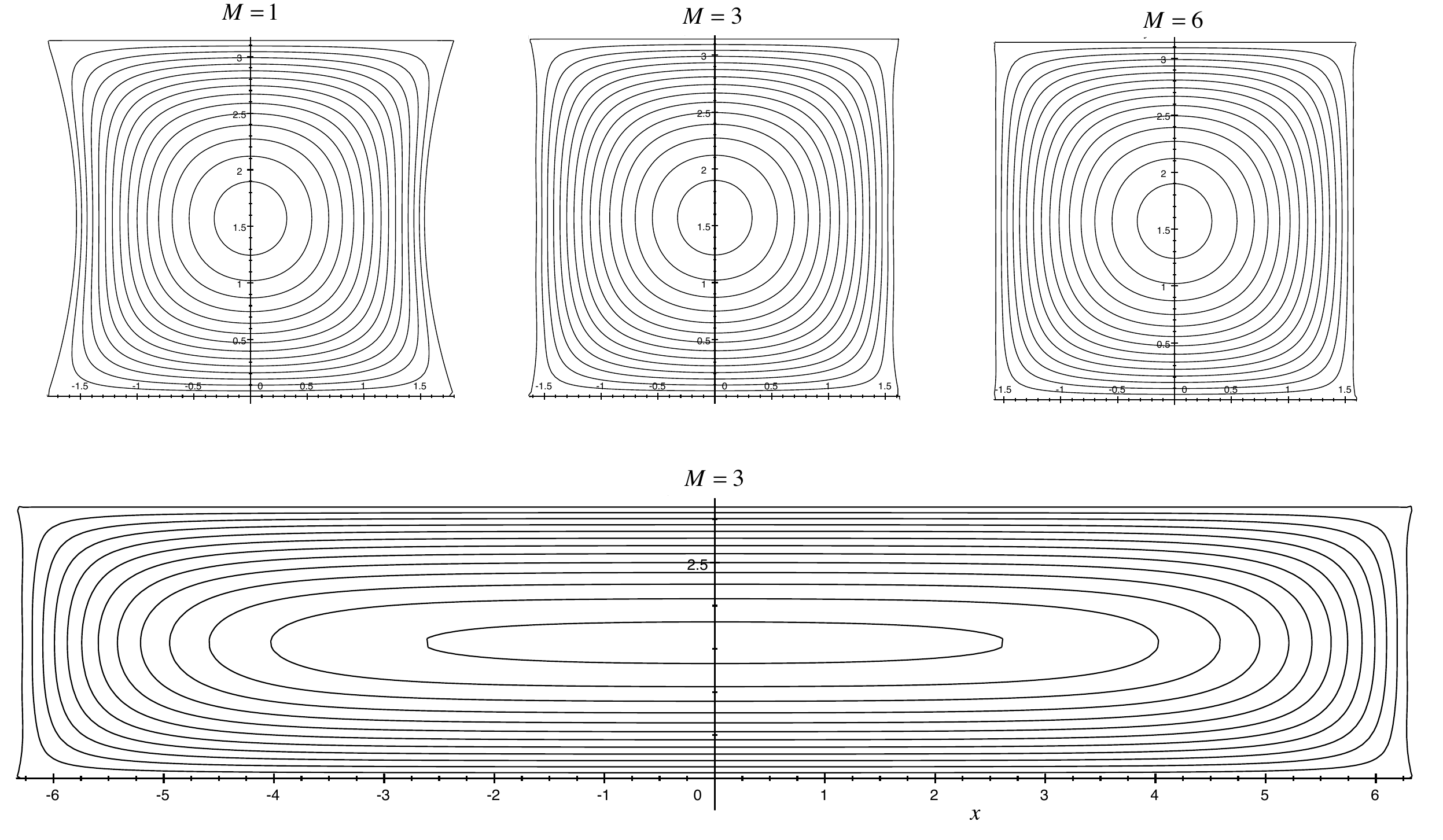}  
\caption{ Exact solutions for near rectangles with $M$ terms in Eq. (\ref{eq:rect}). 
\label{fig:rect}}
\end{center}
\end{figure*}

Because the terms in the expansion are harmonic, however, we can construct exact solutions for  nearly rectangular corners by simply truncating the series, 
\begin{equation}
\tilde{u} =  \frac{\tilde{y} (\pi-\tilde{y})}{2}  -  \frac{4}{\pi}  \sum_{m=1}^M \frac{e^{-(2m-1)\tilde{x}} \sin((2m-1)\tilde{y}) }{(2m-1)^3}
\label{eq:rectcorn}
\end{equation}
since each term in the Fourier series is also harmonic, 
\[
\mbox{Im }e^{-n(\tilde{x}+i\tilde{y})} = - e^{-n\tilde{x}} \sin(n\tilde{y}). 
\]
As shown in Fig.~\ref{fig:rectcorn}(a),(b),(c) for $M=1,2,6$, respectively, the domains corresponding to exact solutions with a few terms are half strips with only slightly warped ends, and the domain with six terms is almost indistinguishable from the rectangular corner.   
In the same way, we can derive exact solutions for finite, nearly rectangular pipes,
\begin{equation}
\tilde{u} =  \frac{1}{2} \tilde{y} (\pi-\tilde{y}) 
-  \frac{4}{\pi}  \sum_{m=1}^M \frac{\cosh((2m-1)\tilde{x}) \sin((2m-1)\tilde{y}) }{(2m-1)^3 \cosh((2m-1)\tilde{L})}
\label{eq:rect}
\end{equation}
with dimensionless height $\pi$ and width $2\tilde{L}$. The Fourier series solution \citep{bruus_book} is again broken into the parabolic Poiseuille flow between flat plates (first term) and a truncated series (second terms) of $M$ harmonic terms, since
\[
\mbox{ Im }\sinh(n(\tilde{x}+i\tilde{y})) = \cosh(n\tilde{x})\sin(n\tilde{y}).
\]
Some examples are shown in Fig. ~\ref{fig:rect} for near squares ($2\tilde{L}=\pi$) and a near rectangle ($2\tilde{L}=4\pi$).  More generally, we can truncate any eigenfunction expansion, or Finite Fourier Transform~\cite{deen_book}, solving Poisson's equation (with homogeneous Robin boundary conditions), since each term must satisfy the homogeneous Laplace's equation.

\section{ Unidirectional flows for given domains } 
\label{sec:geo}

\subsection{ General integral solution }

We have seen that it is easy to generate exact solutions of the form (\ref{eq:gen}) -- but only if we are allowed to choose the boundary geometry.  The difficulty lies in working backwards:  Given a boundary and a convenient choice of the Poiseuille flow or particular solution, $u_p$, what is the harmonic function (shear flow) $u_s$ satisfying the Dirichlet boundary condition (\ref{eq:bc})  that is required to complete the solution (\ref{eq:ups})? 

Using complex analysis, a formal exact solution can be obtained in integral form for any \change{ simply connected domain} as follows, adapting the arguments of \citet{tao1962}. Let $w=f(z)$ be a conformal map from the given $z$ domain to the upper half $w$ plane, \change{ whose existence is guaranteed by the Riemann Mapping Theorem}. The complex potential for the shear flow in  the $w$ plane is then given by the Schwarz integral formula
\begin{equation}
\Phi(w) = \frac{1}{\pi i} \int_{-\infty}^\infty \frac{ u_s(f^{-1}(\xi)) }{\xi - w } d\xi \label{eq:schwarz}
\end{equation}
where $u_s(z)=U(z)-u_p(z)$ is the Dirichlet boundary condition in the $z$ domain.
We thus arrive at the general solution:
\begin{equation}
u(x,y) = u_p(x,y) + \mbox{Re } \Phi(f(x+iy)).  \label{eq:ugeneral}
\end{equation}
which applies to all physically relevant problems,  as in all of our examples, where the flow is bounded by a polynomial function. The reason is that Schwarz integral  (\ref{eq:schwarz}) provides the unique solution to the Dirichlet problem provided only that $\Phi(w)$ is holomorphic and $|w^\alpha \Phi(w)|$ bounded for some $\alpha>0$ in the closed upper half plane.

\subsection{ Rectangular corner }

Let us illustrate the Schwarz integral solution for the case of a rectangular corner considered above. The particular solution in Eq. (\ref{eq:ugeneral}) is again the Poiseuille flow between parallel plates, Eq. ~(\ref{eq:uplate}), which prescribes a parabolic flow profile on the end of the rectangle ($x=0$).  This non-uniform Dirichlet boundary condition makes analytical progress difficult. 

In order to obtain the complex potential for the shear flow, we conformally map the half-strip ($0 < \tilde{y} < i\pi$, $\tilde{x}>0$)  using $\tilde{w}=\cosh(\tilde{z})$ to the upper half plane, where the following Schwarz integral solves the Dirichlet problem,
\begin{equation}
\Phi(\tilde{w}) = \frac{1}{2\pi i} \int_{-1}^{1} \frac{ (\cos^{-1}\xi) (\pi-\cos^{-1}\xi)}{\xi-\tilde{w}} d\xi,
\end{equation}
which can then be mapped back to the $z$ plane to obtain the solution:
\begin{equation}
\Phi(f(\tilde{z})) = \frac{1}{2\pi i } \int_0^\pi \frac{ s (\pi-s) \sin s}{\cos s - \cosh \tilde{z}} ds
\end{equation}
Unfortunately, this integral cannot be expressed in terms of elementary functions, but it can be used as a basis for further analysis, such as asymptotic expansions near the corners where the Fourier series, Eq. (\ref{eq:corn}), loses accuracy.

\begin{figure*}
\begin{center}
\vspace{0.1in}
\includegraphics[width=\linewidth]{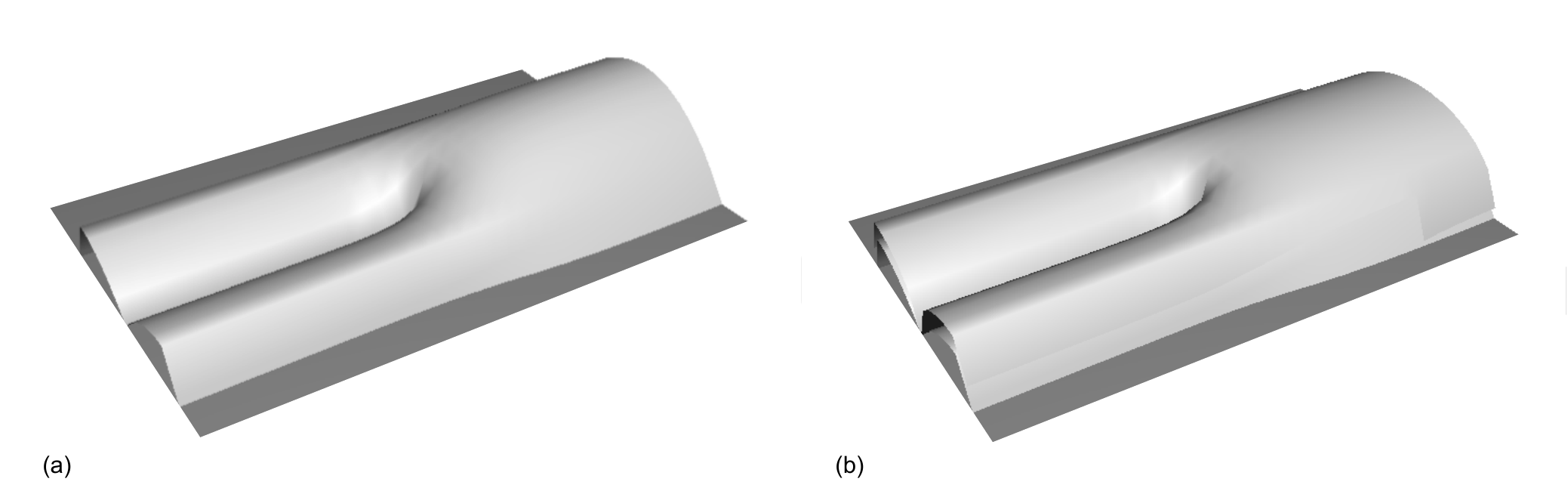}  
\caption{ (a) Exact solution to Poisson's equation for unidirectional flow between flat plates with a parallel half plate inserted at the center, Eqs. (\ref{eq:river1})-(\ref{eq:river2}). (b) Taking a square root, the same solution also describes the  height of the watertable in a strip of land between two rivers or lakes, where another river grows by erosion down the center of the strip.
\label{fig:river}}
\end{center}
\end{figure*}

\subsection{ Symmetric boundaries:  River growth }

The preceding examples illustrate the difficulty of finding exact solutions for given geometries, since the harmonic homogeneous solution usually has non-constant boundary data in order to cancel the particular solution on the boundary.  In some cases, however, the symmetry of the geometry leads at least to piecewise constant boundary data, which is easier to handle.  This is the case when all no-slip ($u=0$) boundaries are iso-velocity contours of the same quadratic-form particular solution in Eq. (\ref{eq:gen}),  {\it i.e.} conic sections. 

Let us consider the following example, which could describe groundwater flow (Fig. ~\ref{fig:analogies}(3)) for the case of river growth in a channel geometry (Fig. \ref{fig:river}).  Recent work has relied on either numerical solutions or the approximation of Poisson's equation by Laplace's equation, thus enabling the more straightforward application of (time-dependent) conformal maps \citep{gubiec2008,cohen2015}.  Here, we formulate a river growth problem that has a simple exact solution for the full model, based on Poisson's equation.  

We solve  $\nabla^2\tilde{u}=-1$ for $\tilde{u} = \kappa h^2 / 2P$, where $h$ is the height of the water table, $P$ is the precipitation rate, and $\kappa$ is the soil permeability. We impose Dirichlet boundary conditions, $\tilde{u}=0$, at $\tilde{y}=\pm 1$ to model a strip of land between two parallel rivers or lakes, and also at $\tilde{y}=0$ for $\tilde{x}<0$, to model a river growing along the center of the strip by erosion. The key simplifying feature is that the central river lies along an iso-velocity line of the particular solution for the height profile in the absence of the river, which is analogous to the Poiseuille flow between parallel plates.  

The solution has the form,
\begin{equation}
\tilde{u}(\tilde{x},\tilde{y}) = 1 - \frac{\tilde{y}^2}{2} + \phi(\tilde{x},\tilde{y})  \label{eq:river1}
\end{equation}
where the harmonic function $\phi(\tilde{x},\tilde{y})$ satisfies $\phi(\tilde{x},\pm1)=0$ and $\phi(\tilde{x}<0,0)=-1$.  The complex potential, $\Phi(\tilde{x}+i\tilde{y})$ can be obtained by conformal mapping:
\begin{equation}
\Phi(z)=\frac{1}{\pi}\log\left( \frac{\sqrt{1-e^{\pi z}} + 1}{\sqrt{1-e^{\pi z}} - 1} \right)  \label{eq:river2}
\end{equation}
where $\phi = \mbox{Im}\Phi(z)$. The solution is shown in Fig. ~\ref{fig:river}.   

The river will grow by erosion along the $x$ axis, according to the Principle of Local Symmetry, at a constant velocity, as a result of translational  invariance. The speed of river growth can be calculated from the solution as   
\begin{equation}
v \sim |\nabla u|^\eta \sim \left|\frac{\partial \tilde{u}}{\partial \tilde{x}}(0,0)\right|^\eta = 2 | \Phi^\prime(0) |^\eta
\end{equation}
in the case of a power-law growth model with exponent $\eta$.  As noted by \citet{cohen2015}, only the harmonic part of the solution determines the gradient at the river tip, leading to growth in the direction of local symmetry.  However,  we see that the harmonic function is related to the particular solution of Poisson's equation, which determines its boundary conditions.  Therefore, the rate of growth down the centerline, as well as the shape of the river as it approaches the centerline (if it were initially displaced to one side), depend on the full solutions of Poisson's equation, including the forcing term. 

In this way, the global patterns formed by ``Poissonian growth" will differ from their approximations by Laplacian growth (Loewner evolution in two dimensions)~\citep{gubiec2008}, in spite of satisfying the same the Principle of Local Symmetry for the tip velocity.   The effect of the forcing term in Poissonian growth is analogous to that of the drift term in transport limited growth:  Local properties, such as the fractal dimension of advection-diffusion-limited aggregation~\citep{bazant2003}, are universal, determined by diffusive Laplacian growth, while the global shape of the pattern is determined by the boundary conditions and the non-Laplacian forcing or drift terms~\citep{davidovitch2005}.  In both cases, diffusion dominates at small length scales because it appears as the highest derivative in the governing transport equation.

\subsection{ \protect{\change{General series solution}} }

\change{
In this article, we have focused on closed-form exact solutions for the flow profile, which provide many insights, but they are not convenient for analytically evaluating integrals of the solution, such as the total flow rate or hydraulic resistance, as a result of either implicit boundary representations (Section \ref{sec:implicit}) or intractability of the Schwarz integral (\ref{eq:schwarz}).  On the other hand, it is possible to derive general series  solutions for the flow profile and hydraulic resistance in terms of the Maclaurin series coefficients of the conformal map from the unit disk to the domain, as first discovered by R. M. Morris in 1939 \cite{morris1939}.  The reader is referred to excellent book of \citet{polya_book} for a discussion of complex series expansions, as well as rigorous bounds, for the ``torsional rigidity" of a prismatic beam, which is analogous to the hydraulic conductance of a pipe.  This connection has recently been exploited by \citet{kacimov2002} to calculate the permeability of cylinder packings, but otherwise the general solution has largely escaped notice in hydrodynamics.  Other analogies from Fig.~\ref{fig:analogies} include the average temperature of an electrical wire, the mean escape time from a cylinder,  and the average height of a soap bubble or draining aquifer.
}

\section{ Unidirectional flow with non-constant forcing }

\subsection{ General solution of Poisson's equation in two dimensions }
The same mathematical methods above can be extended to solve the general Poisson's equation,
\begin{equation}
-\nabla^2 u = \rho,   \label{eq:pois}
\end{equation}
for a non-constant forcing term, $\rho(x,y)$ (charge density).   Although this situation no longer describes unidirectional flow or soap bubbles, it is relevant for the many other physical problems sketched in Fig. \ref{fig:analogies}.  Besides electrostatics in a non-uniformly  charged cylinder, important applications include resistive heating in a wire with nonuniform current density, reaction-diffusion in a catalyst rod with non-uniform reaction rates, and electrokinetic phenomena in a typical pore or microchannel with non-uniform diffuse charge density.

As long as a particular solution $u_p$ can be derived for a given $\rho$, then families of exact solutions can be generated by simply adding harmonic functions $u_s$, which solve the homogeneous (Laplace's) equation.  This idea can be exploited in any number of dimensions, but it is easiest to generate harmonic functions in two dimensions as the real parts of analytic functions, yielding the general solution, Eq. (\ref{eq:ugeneral}).  For a given geometry and particular solution, the harmonic function is uniquely determined by the Schwarz integral, Eq. (\ref{eq:schwarz}), in terms of the conformal map, $w = f(z)$, of the domain to the upper half plane. 

For example, in the case of a general $N$th-order polynomial charge density in two dimensions,
\begin{equation}
\rho(x,y) = \sum_{m=0}^N \sum_{n=0}^N \rho_{m,n} x^m y^n,
\end{equation}
the particular solutions of Eq. (\ref{eq:pois}) are $(N+2)$-order polynomials,
\begin{equation}
u_p(x,y) = \sum_{m=0}^{N+2} \sum_{n=0}^{N+2} A_{m,n} x^m y^n
\end{equation}
whose coefficients are subject to the constraints,
\begin{equation}
(m+2)(m+1)A_{m+2,n} + (n+2)(n+1)A_{m,n+2} = - \rho _{m,n} 
\end{equation}
and
\begin{equation}
A_{m+2,N+1}=A_{m+2,N+2}=A_{N+1,n+2}=A_{N+2,n+2}=0
\end{equation}
for all $m,n\in \{0,1,...,N\}$.   In the case of constant charge density, $N=0$, this general polynomial solution reduces to the quadratic-form solution for a conic section domain, Eqs. (\ref{eq:conic})-(\ref{eq:conicA}).

\subsection{ Unidirectional linear electrokinetic phenomena }

Consider a long straight pore or microchannel of arbitrary cross section, filled with a liquid electrolyte, having an arbitrary surface charge distribution $q_s(x,y)$ that does not vary with axial position.  In equilibrium, there is a non-uniform diffuse charge density, $\rho_{eq}(\phi)$, resulting from imbalances in ionic concentrations, which depends on the electrostatic potential, $\phi(x,y)$, determined self-consistenly by Poisson's equation,
\begin{equation}
-\varepsilon\nabla^2\phi = \rho_{eq}(\phi)    \label{eq:poi}
\end{equation}
and the electrostatic boundary condition, $-\varepsilon\ \hat{n}\cdot\nabla \phi=q_s$. 
The zeta potential, $\zeta(x,y)$, is defined as the value of $\phi(x,y)$ at the surface.

\begin{figure*}
\begin{center}
\vspace{0.1in}
\includegraphics[width=5in]{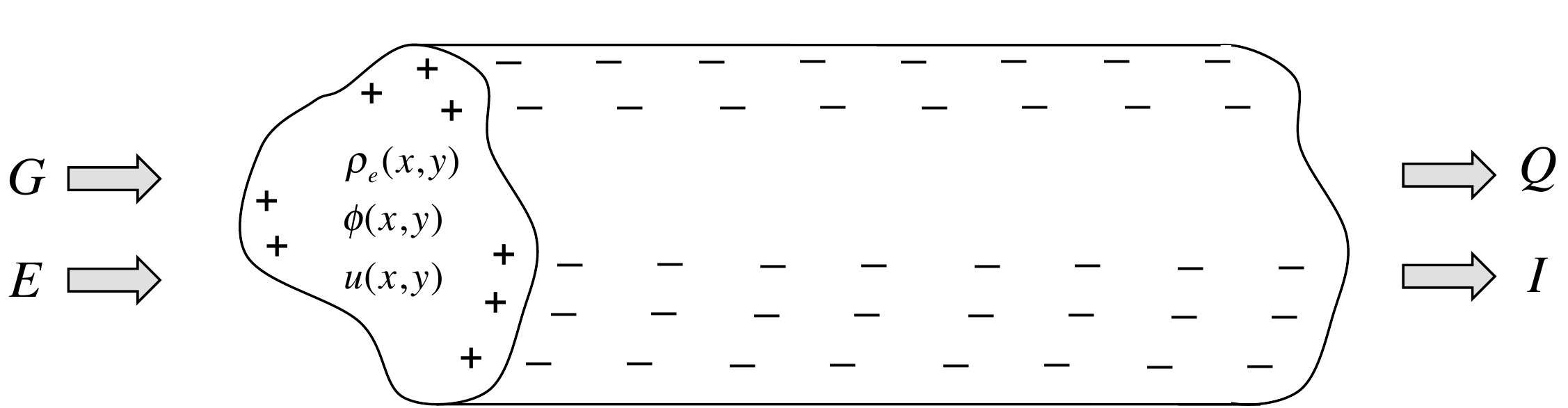}  
\caption{ Unidirectional electrokinetic phenomena for a liquid electrolyte in a straight nanochannel or pore having irregular cross section and non-uniform surface charge density.  In the absence of applied concentration gradients,  the Onsager linear response matrix in Eq. (\ref{eq:Lij}) relates two fluxes, the flow rate $Q$ and electrical current $I$, to two forces, the axial pressure gradient $G$ and electric field $E$.
\label{fig:EK}}
\end{center}
\end{figure*}

In linear response to a small axial electric field, $E$, there is a unidirectional electro-osmotic flow with axial velocity, $u_e(x,y)$, given by the balance of viscous stress and electrostatic body force (Maxwell stress) on the fluid,
\begin{equation}
-\mu\nabla^2 u_e = \rho_{eq}(\phi)E    \label{eq:ue}
\end{equation}
There can also be a unidirectional pressure-driven flow, $u_g(x,y)$, in response to an axial pressure gradient, $G$,
\begin{equation}
-\mu\nabla^2 u_g=G   \label{eq:ug}
\end{equation}
which results in an axial streaming current (advection of diffuse charge). Each of the flow fields satisfies no-slip on the boundary, $u_e=u_g=0$.  

These linear electrokinetic phenomena, sketched in Fig. ~\ref{fig:analogies}(14)-(15), are thus described by three Poisson equations, Eqs. (\ref{eq:poi})-(\ref{eq:ug}), two of which which have non-constant (and nonlinear) forcing terms, related to the nonuniform distribution of diffuse charge.  The general solution above could be applied to obtain the flow and current for a given distribution $\rho_{eq}(x,y)$, or it could be used as the basis for a numerical solution scheme.  Fortuitously, we do not need to solve the equations to obtain some useful relationships by taking advantage of the fact that each unknown function can be expressed as the sum of a particular solution and a harmonic homogeneous solution. By combining Eqs. (\ref{eq:poi}) and (\ref{eq:ue}), the electro-osmotic flow can be expressed as 
\begin{equation}
u_e = \frac{\varepsilon(\phi-\psi)E}{\mu}   \label{eq:uepsi}
\end{equation}
where the first term is a particular solution and the second is the homogeneous solution, satisfying Laplace's equation
\begin{equation}
\nabla^2\psi=0  \label{eq:harm}
\end{equation}
The no-slip condition implies that $\psi(x,y)=\zeta(x,y)$ on the boundary.  In the limits of uniform zeta potential and thin double layers, where $\rho_e\to 0$ and $\phi\to 0$ over the central region of the pore,  we recover  the classical Helmholtz-Smoluchowski slip formula, 
\begin{equation} 
u_e \to - \frac{\varepsilon \zeta E}{\eta}
\end{equation}
for electro-osmotic plug flow.

\subsection{ Proof of Onsager reciprocity  }

\citet{onsager1931a,onsager1931b} argued that the linear response matrix, which relates thermodynamic forces and fluxes near equilibrium, must be symmetric, based on the reversibility of the microscopic equations of motion. Onsager reciprocity is widely accepted, even in situations where it has not been rigorously proved from the equations of motion, and has become a fundamental postulate of non-equilibrium thermodynamics ~\citep{degroot_book}. In particular, the linear response matrix for electrokinetic phenomena is always assumed to be symmetric \citep{hunter_book}, even for anisotropic  particles \citep{long1998}, channels \citep{ajdari2001}, and surfaces \citep{bahga2010}.  In the case of a circular cylindrical capillary, reciprocal relations have been derived from the fundamental transport equations by \citet{gross1968} and \citet{peters2016}, also allowing for linear response to  salt concentration gradients.   For general porous media, formal proofs of Onsager reciprocity for linear electrokinetics are also available, based on the assumptions of homogenization theory  \citep{looker2006,schmuck2015}. 
Here, we simply exploit the mathematics of unidirectional flow. 

The preceding results can be used to derive electrokinetic reciprocal relations for a straight pore of arbitrary cross-sectional shape and non-uniform surface charge around its perimeter, sketched in Fig. ~\ref{fig:EK}.   There is no constraint on the topology of the cross section.  A non-simply-connected domain would describe a medium with parallel pores, where the fluxes and flows are unidirectional.   

Neglecting salt concentration gradients, the total flow rate $Q$ and total current $I$ are linearly related to the applied electric field and pressure gradient via the electrokinetic conductance matrix,
\begin{equation}
\left( \begin{array}{c} Q \\ I \end{array} \right)
 = \left( \begin{array}{cc} L_H & L_{EO} \\ L_{SC}  & L_E \end{array} \right)
 \left( \begin{array}{c} G \\ E \end{array} \right)   \label{eq:Lij}
\end{equation}
where $L_H$, $L_E$, $L_{EO}$ and $L_{SC}$ are the hydraulic (Darcy), electrical (Ohmic), electro-osmotic and streaming-current conductances, respectively.   The goal is to prove $L_{SC} = L_{EO}$.

\vspace{0.1in} 

{\bf Proof:}  
\begin{eqnarray*}
L_{SC} &=& \frac{1}{G} \iint \rho_e u_g \; dxdy \\
&=& - \frac{\varepsilon}{G} \iint (\nabla^2 \phi) u_g \; dxdy \ \ \ \mbox{ (Eq. (\ref{eq:poi}))} \\
&=& - \frac{\varepsilon}{G} \iint (\nabla^2 (\phi-\psi)) u_g \; dxdy \ \ \ \mbox{ (Eq. (\ref{eq:harm}))} \\
&=& - \frac{\varepsilon}{G} \iint (\phi - \psi) \nabla^2 u_g \; dxdy \ \ \ \mbox{ (Eq. (\ref{eq:green}))} \\
&=&  \frac{\varepsilon}{\mu} \iint (\phi - \psi)  \; dxdy \ \ \ \mbox{ (Eq. (\ref{eq:ug}))} \\
&=&  \frac{1}{E} \iint u_e  \; dxdy \ \ \ \mbox{ (Eq. (\ref{eq:uepsi}))} \\
&=& L_{EO}
\end{eqnarray*}
In the key step above, we perform two integrations by parts (Green's second identity),
\begin{eqnarray}
\iint (\nabla^2 f_1) f_2 \; dxdy &=& \oint \hat{n}\cdot(f_2 \nabla f_1 - f_1 \nabla f_2) \nonumber \\ 
& & + \iint f_1 (\nabla^2 f_2) \; dxdy \label{eq:green}
\end{eqnarray}
where the line integral vanishes since $f_1 = \phi-\psi=0$  and $f_2 = u_g = 0$ on the surface.

\section{ Conclusion } 

This paper provides an overview of exact solutions  and physical analogies for unidirectional flow.  Through a variety of examples, we have shown how to construct approximate geometries that admit simple exact solutions by  manipulating the harmonic homogeneous solution to Poisson's equation.  We have identified seventeen physical analogies for unidirectional flow, including many that may not have been noted before, in geological flows, electrokinetics, electrochemistry, and stochastic processes.  In these fields, the classical mathematics of unidirectional flow may find some new applications.

\change{
Here, we have focused on exact solutions for the flow profile, but applications often require only its area integral, the hydraulic conductance, or analogous quantities.  There is a vast body of mathematical results from the theory of elasticity that could be exploited, not only from complex analysis but also from geometry, based on so-called ``isoperimetric inequalities"~\cite{polya_book,payne1967}.  It is beyond the scope of this paper to apply these ideas to Poiseuille flow and its many analogies, so this task will be left for a forthcoming publication.  For now, we close with an isoperimetric inequality proposed by A. Acrivos:  
\begin{quote}
{\it A circular pipe has the least hydraulic resistance for a given cross-sectional perimeter. }\\
\end{quote}
We have not found this theorem stated in the literature, although its proof follows from the general theory~\cite{polya_book}. 
In his seminal paper, \citet{venant1855} made a similar conjecture, that a circular beam has the greatest torsional rigidity for a given cross-sectional area.  Almost a century passed before a proof was provided by G. P{\' o}lya  \cite{polya1948}, who laid the groundwork to understand the shape dependence of solutions to Poisson's equation. 
}

\section*{ Acknowledgments } 
This work was partially supported by the Global Climate and Energy Project at Stanford University and by the US Department of Energy, Basic Energy Sciences through the SUNCAT Center for Interface Science and Catalysis.

\bibliography{uni}

\end{document}